\documentclass[notitlepage,floats,aps,nofootinbib,preprintnumbers,superscriptaddress,twocolumn,prd]{revtex4-1}

\usepackage{amsmath,amssymb,amsfonts}
\usepackage{graphicx}
\usepackage{xcolor}
\usepackage{mathrsfs}
\usepackage{slashed}
\usepackage[utf8x]{inputenc}
\usepackage{hyperref}
\usepackage{hyperref}
\hypersetup{colorlinks,citecolor=blue,urlcolor=blue,linkcolor=blue}
\usepackage{subfigure}
\usepackage{mdframed}
\usepackage{ulem}
\usepackage{units}
\usepackage[english]{babel}

\def\beq{\begin{equation}\begin{aligned}}
\def\eeq{\end{aligned}\end{equation}}

\newcommand{\mb}[1]{\boldsymbol{#1}}
 \newcommand{\rep}[1]{\mathbf{#1}}
\newcommand{\conjrep}[1]{\overline{\mathbf{#1}}}

\begin{document}
\title{A QCD R-Axion}

\author{James Unwin}
\affiliation{Department of Physics, University of Illinois at Chicago, Chicago, IL 60607, USA}
\author{Tom Yildirim}
\affiliation{Department of Physics, Keble Road, University of Oxford,  OX1 3RH, UK}
\begin{abstract} 
R-parity can be extended to a continuous global U(1)${}_R$ symmetry. We investigate whether an anomalous U(1)${}_R$ can be identified as the PQ symmetry suitable for solving the strong CP problem within supersymmetric extensions of the Standard Model.  In this case, U(1)${}_R$ is broken at some intermediate scale and the QCD axion is the R-axion. Moreover, the R-symmetry can potentially be gauged via the Green-Schwarz mechanism within completions to supergravity, in order to evade the axion quality problem. Obstacles to realizing this scenario are highlighted and phenomenologically viable approaches are identified.
\vspace*{1mm}
\begin{center}
    {\textit{Dedicated to the memory of Professor Fidel A.~Schaposnik.}}
\end{center}
\end{abstract}

\maketitle

\newpage
\section{Introduction}
\label{sec:Introduction}

A spontaneously broken anomalous global U(1) symmetry is a key ingredient of the Peccei-Quinn (PQ) resolution \cite{Peccei:1977hh,Peccei:1977ur} of the strong CP Problem. This new global symmetry is called the PQ symmetry. Supersymmetric extensions of the Standard Model also motivate a global symmetry U(1)${}_R$, the ``R-symmetry'' \cite{Fayet:1974pd,Salam:1974xa}, which can be associated to rotations of the superspace coordinates $\{\vartheta,\bar\vartheta\}\rightarrow \{e^{-i\lambda}\vartheta,e^{i\lambda}\bar\vartheta\}$. 
Famously, the Minimally Supersymmetric Standard Model (MSSM) is formulated with R-parity, since this avoids fast proton decay \cite{Farrar:1978xj,Dimopoulos:1981zb}, which can be viewed as a $Z_2$ subgroup of U(1)${}_R$. It is natural to ask if one can construct R-symmetric models in which U(1)${}_R$ is anomalous under QCD and spontaneously broken in such a manner that it may be utilised to implement the PQ mechanism while avoiding observational exclusion. In this paper, we present a class of models as \textit{proof of principle} that, indeed, the PQ symmetry could be identified with U(1)${}_R$.

The spontaneous breaking of an approximate global symmetry leads to a pseudo Nambo-Goldstone Boson (pNGB), and U(1)${}_R$ is no different in this respect, leading to a pNGB called the R-axion \cite{Farrar:1982te,Fayet:1980ad,Fayet:1981,Bagger:1994hh}. The prospect that the R-axion could be related to the Strong CP Problem has previously been largely\footnote{In the final stages of this work a paper by Dvali, Kobakhidze and Sakhelashvili \cite{Dvali:2024dlb}, appeared which also discussed the prospect of linking the R-axion to the strong CP problem, taking a different perspective.} neglected, or rejected on phenomenological grounds, see e.g.~\cite{Banks:1993en,Nelson:1995hf,Dobrescu:2000yn,Miller:2003hm,Carpenter:2009sw}.
There are several difficulties in identifying the PQ symmetry with U(1)${}_R$, in particular, one must have R-charged states with appropriate coupling structures to allow one to rotate away the QCD angle $\bar\theta$, obtain phenomenologically viable masses for the gauginos and axion, and avoid explicit violations of R-symmetry from constant superpotential terms or Planck scale operators. 
We have found these obstacles can be navigated more readily if one allows the MSSM superfields to have general R-charges constrained only by phenomenological requirements, departing from a common assumption in the literature.  In this work, we construct several scenarios in which the U(1)${}_R$ symmetry is identified with the anomalous symmetry which implements the PQ mechanism. 
Notably, the PQ symmetry is arguably \textit{ad hoc} in many implementations, and identifying the PQ symmetry with U(1)${}_R$ presents an elegant economy.

This paper is structured as follows; in Section \ref{sec:2} we start by outlining the restriction on R-charge assignments that allow one to construct Yukuawa terms and the Weinberg operator. We then highlight a mechanism that links the $\mu$-term  to the R-breaking scale. 
 In Section \ref{sec:3}, we discuss issues relating to explicit violations of U(1)${}_R$ connected to tuning the cosmological constant and highlight scenarios that circumvent this problem by tuning the scalar potential to zero without adding a constant in the superpotential.
 Section \ref{sec:4} examines phenomenological aspects, in particular, manners to obtain viable gaugino masses,  and requirements for avoiding fast proton decay.
Section \ref{sec:5} discusses the UV origins of the R-symmetry, highlighting that the global U(1)${}_R$ can potentially be gauged via the Green-Schwarz mechanism within  supergravity, thus evading the axion quality problem. 
  We provide a summary in Section \ref{sec:con}, alongside some remarks regarding the implication for cosmology and experimental searches.

\vspace{-3mm}
\section{Identifying  U(1)\texorpdfstring{${}_R$}{{}_R} with  the PQ symmetry}
\label{sec:2}

Under the U(1)${}_R$ symmetry the scalar components of chiral superfields transform as $\phi\rightarrow e^{i\lambda R}\phi$ whereas fermion components transform $\chi\rightarrow e^{i\lambda (R-1)}\chi$. In particular,  R-transformations on the fermion component act as axial U(1) rotations with charge $R-1$ \cite{Fayet:1974pd,Salam:1974xa}. The typical approach in R-symmetric supersymmetric extensions of the Standard Model is to assign Standard Model superfields R-charges as follows  \cite{Hall:1990hq}
\beq
R[\mb{Q}]&=R[\mb{U^c}]=R[\mb{D^c}]=R[\mb{L}]=R[\mb{E^c}]=1,\\
 R[\mb{H_u}]& =R[\mb{H_d}]= 0,
\label{binary}
\eeq
the superspace coordinate $\vartheta$ canonically has unit R-charge. These assignments are motivated by simplicity and comparisons to R-parity; however, nothing requires the MSSM fields to carry only binary R-charges, and in what follows, we will explore generalized R-charge assignments.
  
With the charge assignments of eq.~(\ref{binary}), the MSSM superpotential terms are each invariant under U(1)${}_R$ (i.e.~with net R-charge 2 which cancels against the measure ${\rm d}^2\vartheta$), with the exception of the $\mu$-term ($\mu \mb{H_uH_d}$) which has zero net R-charge. This implies that the MSSM superpotential is R-symmetric provided $\mu=0$. The $\mu$-term can be subsequently introduced through R-symmetry breaking terms. Notably, supersymmetric extensions of the Standard Model with global U(1)${}_R$ invariance which is exact at the weak scale have been explored in e.g.~\cite{Farrar:1982te,Hall:1990hq,Fayet:1975yi,Fayet:1976et,Fayet:1977yc,Fox:2002bu,Kribs:2007ac}. In contrast, here we propose to spontaneously break the R-symmetry at some intermediate scale in order to successfully implement the PQ mechanism.

\subsection{R-Symmetric Superpotentials}

 We start by parameterising the R-charge assignments of  MSSM fields and identifying the restrictions needed in order to reproduce the structure and successes of the MSSM at low energy.  After fixing $R[\vartheta]=1$ there is no freedom with the R-charges of the gravitino \cite{Freedman:1976uk}  and gauginos, which are required to carry unit R-charge. 
To proceed, let us parameterise the net R-charge of the Higgs superfield bilinear as follows $R[\mb{H_uH_d}]=A$ and write 
\beq
&R[\mb{H_u}]=A\cos^2\beta,\\
&R[\mb{H_d}]=A\sin^2\beta.
\label{eqA}
\eeq
As we discuss in the next subsection the R-symmetry forbids the $\mu$-term provided that  $A\neq2$.

Importantly, we should choose the R-charges for the MSSM fields such that the Yukawa terms are invariant. 
There is a modest degree of parameter freedom, so we choose to parameterise $R[\mb{Q}]=Q$, where $Q$ can be identified as the generator of Baryon number. Considering, for instance, the operator $\mb{H_uQU^c}$ we require that
\beq
R[\mb{H_uQU^c}]=2 ~~ \Rightarrow ~~  R[\mb{U^c}] =2 - A\cos^2\beta - Q~.
\eeq
Similarly, for the other Yukawa terms. Additionally, we would like $R[\mb{LH_uLH_u}]=2$ to allow for neutrino masses, thus we require
\beq 
R[\mb{LH_u}]=1 \quad  \Rightarrow \quad R[\mb{L}]=1-A\cos^2\beta~.
\eeq
We collect together the relations between R-charges of various superfields in the ``U(1)${}_R$'' column of Table \ref{Tab1}.  For $A=0$ and $Q=1$, one arrives at the binary R-charge assignments of eq.~(\ref{binary}). Moreover, we highlight that for $R[\mb{f}]\neq1$ (for $\mb{f}=\mb{Q},\mb{U^c},\mb{D^c},\mb{L},\mb{E^c}$)  the Standard Model fermion associated to $\mb{f}$ will carry non-zero R-charge.  

Interestingly, it has been argued  that rationality of R-charges is not required since in certain UV completions into supergravity U(1)${}_R$ is not embedded into a larger group, nor is there a condition for R-charge quantization~\cite{Castano:1995ci}.
However, it is certainly aesthetic (and familiar) to have rational charges and one can find such cases with a careful restriction on the R-charges, for instance, in the ``U$(1)_R |_{A}$'' column we fix $A=1/(3\cos^2\beta)$, and in the ``U$(1)_R |_{A,\beta}$''  column we further fix $\tan\beta=3$ to arrive at a simple set of rational R-charges.

\begin{table}[t]
\begin{center}
\def\str{\vrule height15pt width0pt depth7pt}
\begin{tabular}{| c | l | c | c | c | c |  c|}
    \hline\str
    ~Field ~&~ Gauge rep. ~&~ U(1)${}_R$  &${\rm U}(1)_R |_{A}$ &${\rm U}(1)_R |_{A,\beta}$\\
    \hline\str
    ~~$\mb{Q}$ & ~~$\,(\rep{3},\rep{2})_{1/6}$ & $Q$ & $Q$ & $Q$\\ 
    \hline\str
    ~~$\mb{U^c}$ & ~~$\,(\conjrep{3},\rep{1})_{-2/3}$ & $2-Q-Ac^2_\beta$  & $5/3-Q$ & $5/3-Q$ \\ 
    \hline\str
    ~~$\mb{D^c}$ & ~~$\,(\conjrep{3},\rep{1})_{1/3}$ & $2-Q-As^2_\beta$  & $2-Q-t_\beta^2/3$ & $-1-Q$  \\ 
    \hline\str
    ~~$\mb{L}$ & ~~$\,(\rep{1},\rep{2})_{-1/2}$ & $1-Ac^2_\beta$   & $2/3$ & $2/3$   \\ 
    \hline\str
    ~~$\mb{E^c}$ & ~~$\,(\rep{1},\rep{1})_1$ & $1+A(c^2_\beta-s^2_\beta)$ & $4/3-t^2_\beta/3$ &  $-5/3$ \\ 
    \hline\str
    ~~$\mb{H_u}$ & ~~$\,(\rep{1},\rep{2})_{1/2}$ & $Ac^2_\beta$  & 1/3 & 1/3  \\ 
    \hline\str
    ~~$\mb{H_d}$ & ~~$\,(\rep{1},\rep{2})_{-1/2}$ & $As^2_\beta$  & $t^2_\beta/3$ & $3$  \\ 
            \hline
\end{tabular}
\vspace{4mm}\caption{
General R-symmetry assignments for the MSSM fields, restricting only by the requirement that these permit the Yukawa and Weinberg operators. We use the shorthand $c_\beta=\cos\beta$, $s_\beta=\sin\beta$, $t_\beta=\tan\beta$. In the U$(1)_R |_{A}$ column we fix $A=1/(3\cos^2\beta)$, and in the U$(1)_R |_{A,\beta}$  we further fix $\tan\beta=3$ to arrive at a simple set of charges. Additional field content is given in Table \ref{Tab2}.
}
\label{Tab1}
\end{center}
\end{table}

\subsection{U(1)${}_R$ and the $\mu$-problem}
\label{2.2}
The presence of the term $\mu \mb{H_u H_d}$ in the MSSM superpotential has long been regarded as problematic since one requires $\mu\lesssim$ TeV to retain a   resolution to the hierarchy problem, whereas naturalness arguments naively imply  $\mu\sim M_{\rm Pl}$. The $\mu$-problem can be resolved by forbidding the term in the superpotential and then generating an effective $\mu$-term through the Giudice-Masiero mechanism \cite{Giudice:1988yz},  Kim-Nilles mechanism \cite{Kim}, or similar. Notably, in R-symmetric supersymmetric extensions of the Standard Model one does not need to introduce a new \textit{ad hoc} symmetry to remove the $\mu$-term, as it is achieved via the U(1)${}_R$ symmetry for appropriate charge assignments for ${\mb H_u}$ and ${\mb H_d}$.
With $R[\mb{H_u H_d}]\neq2$,  the $\mu$ term is forbidden in the renormalizable superpotential. As alluded to above, there are a number of mechanisms through which the $\mu$-term can be reintroduced. 

We shall emulate the approaches recently re-examined in Bhattiprolu \& Martin \cite{Bhattiprolu:2021rrj} (building on Murayama-Suzuki-Yanagida \cite{Murayama}), which considered superpotentials involving additional Standard Model singlet chiral superfields charged under the PQ symmetry. These models economically link the PQ breaking scale with the generation of the $\mu$-term. 
Here we consider one possible construction of this type and, in contrast to \cite{Bhattiprolu:2021rrj}, we assume that the spontaneously broken symmetry is U(1)${}_R$ rather than an \textit{ad hoc} PQ symmetry.  
Specifically, we consider the following superpotential contribution \cite{Murayama,Bhattiprolu:2021rrj}.
\beq
\label{Hitoshi}
W_{\rm \mu}=\frac{1}{\Lambda}\left(\mb{XYH_uH_d}+\mb{X^3Y}\right)~,
\eeq
where $\Lambda$ has mass dimension one. It what follows typically, we shall identify $\Lambda=M_{\rm Pl}$.  Introducing two fields $X$ and $Y$  is useful for constructing a stabilised potential for the additional scalars (via the $X^3Y$ term).

 We note that our models are distinct from earlier scenarios of \cite{Murayama,Bhattiprolu:2021rrj} since our symmetry is an R-symmetry. 
Importantly, this changes the requirements on the allowed superpotential terms since now they should carry net R-charge 2 to cancel against the $\mathrm{d}\vartheta$.
Thus for eq.~(\ref{Hitoshi}) to be invariant under U(1)${}_R$ we require
\beq
R[\mb{XYH_uH_d}]=R[\mb{X^3Y}]=2~.
\eeq
The latter restriction implies
\beq
3R[\mb{X}]+R[\mb{Y}]=2 \quad \Rightarrow \quad R[\mb{X}]=\frac{1}{3}(2-R[\mb{Y}])~.
\eeq
Recall that in the above we have made the parameterisations for the Higgses
$R[\mb{H_u}]=A\cos^2\beta$ and
$R[\mb{H_d}]=A\sin^2\beta$
(with $A\neq2$ in order to forbid the $\mu$-term). Requiring that the first operator in $W_\mu$ is invariant under U(1)${}_R$ places a further restriction on the R-charges
\beq
R[\mb{XYH_uH_d}]&=A+\frac{2}{3}(1+R[\mb{Y}])=2~,
\label{eq5}
\eeq
and it follows that
\beq
R[\mb{Y}]&=2-\frac{3A}{2}~,\\
R[\mb{X}]&=\frac{A}{2}~.
\label{eq7}\eeq
For easy reference, we summarise these requirements in Table \ref{Tab2} (supplementing Table \ref{Tab1}).

\begin{table}[t]
\begin{center}
\def\str{\vrule height15pt width0pt depth7pt}
\begin{tabular}{| c | l | c | c | c | c |  c|}
    \hline\str
    ~Field ~&~ Gauge rep. ~&~ U(1)${}_R$  &${\rm U}(1)_R |_{A}$ &${\rm U}(1)_R |_{A,\beta}$  \\
    \hline\str
    ~~$\mb{X}$ & ~~$\,(\rep{1},\rep{1})_{0}$ & $A/2$  & $1/(6c^2_\beta)$ & $5/3$ \\
            \hline\str
    ~~$\mb{Y}$ & ~~$\,(\rep{1},\rep{1})_{0}$ & $2-3A/2$   & $2-1/(2c^2_\beta)$ & $-3$  \\
        \hline\hline\str
    ~~$\mb{\Phi_g}$ & ~~$\,(\rep{8},\rep{1})_{0}$ & $0$ & $0$ & $0$  \\
        \hline\str
    ~~$\mb{\Phi_W}$ & ~~$\,(\rep{1},\rep{3})_{0}$ & $0$ & $0$ & $0$ 
   \\ \hline
\end{tabular}
\vspace{3mm}\caption{Supplementary field content in the viable model constructed here. This table follows the same format as Table \ref{Tab1}.  The first set of two rows provides the Standard Model singlet chiral superfields responsible for spontaneously breaking the R-symmetry, see Section \ref{2.2}. The latter set of three rows gives the charge assignments of the gauge adjoint superfields which allow for Dirac gaugino masses, see Section \ref{3.1}. The  adjoint superfield 
$\mb{\Phi_W}$ is not strictly needed if one is comfortable with heavier MSSM superpartners. The field content and charge assignments are not unique.}

\label{Tab2}
\end{center}
\end{table}

The U(1) symmetry will simplify the superpotential, forbidding $\mb{XH_uH_d}$ and $\mb{YH_uH_d}$, in the case that $A\neq0,4/3$.  Further, for $A\neq1$ the U(1) forbids $\mb{H_uH_dH_uH_d}$, $\mb{X^2H_uH_d}$, and $\mb{Y^2H_uH_d}$ in the superpotential. While there is nothing immediately problematic with including any or all of these five terms, our aim is to recover the simple form of the superpotential $W_{\mu}$ in eq.~(\ref{Hitoshi}). Additionally, to avoid tadpoles we also require that $R[\mb{X}],R[\mb{Y}]\neq2$. Thus to make the connection with \cite{Murayama,Bhattiprolu:2021rrj} we shall make the minor restriction: 
 \beq
\label{0124}
A\neq0,~1,~\frac{4}{3},~2,~4. 
\eeq 
The superpotential contribution of eq.~(\ref{Hitoshi}) generates Higgsino masses once the $X$ and $Y$ scalar components obtain VEVs, as can be seen from the soft supersymmetry breaking terms 
\beq
\mathcal{L}_{\rm soft}=(aXYH_uH_d+bX^3Y +{\rm c.c.})-m_X^2|X|^2-m_Y^2|Y|^2~,
\eeq
where $m_X,m_Y\sim m_{\rm soft}$ and  the coupling $a$ and $b$ are dimensionless parameters of order
\beq
a,b\sim\frac{m_{\rm soft}}{\Lambda}~.
\eeq
As shown in \cite{Murayama,Bhattiprolu:2021rrj} if $m_X^2$ and $m_Y^2$ are negative or sufficiently small, then the potential has a local minimum for $X$,$Y$ VEVs of order
\beq
\label{<X>}
\langle X\rangle\sim \langle Y\rangle\sim\sqrt{m_{\rm soft}\Lambda}~.
\eeq
This is appropriate to generate a $\mu$ term at the weak scale 
\beq
\mu \sim\frac{\langle X\rangle \langle Y\rangle}{\Lambda}\sim m_{\rm soft}.
\eeq

We highlight that the Kim-Nilles mechanism implementation for reintroducing the $\mu$ term and linking it to the soft masses via the superpotential $W_{\mu}$ given in eq.~(\ref{Hitoshi}) is one of several possible simple variants, as recently noted in \cite{Bhattiprolu:2021rrj}. For instance, one could alternatively consider $W\supset \mb{X^2 H_uH_d}+\mb{X^2Y^2}$ in which case the mechanism would be unaltered in spirit, but the details and charge assignments of $\mb X$ and $\mb Y$ would be different. We provide a brief discussion on these variant Kim-Nilles implementations in Appendix \ref{ApB}. Furthermore, one could explore generating the $\mu$-term via K\"ahler terms \cite{Giudice:1988yz} which would also alter the phenomenological details.

\subsection{Spontaneous R-breaking}

Once the R-charged scalars develop non-zero VEVs the global U$(1)_R$ is spontaneously broken. 
Thus the VEVs of the $X$, $Y$ scalars define the scale of R-symmetry breaking, which in this case is also the scale of PQ breaking, given by
\beq
f_a\sim\sqrt{R[X]^2\langle X\rangle^2+R[Y]^2\langle Y\rangle^2}~.
\eeq
By inspection of eq.~(\ref{<X>}), for $\Lambda=M_{\rm Pl}$ and low scale supersymmetry $m_{\rm soft}\sim1$ TeV one has 
\beq
f_a\sim\sqrt{m_{\rm soft}M_{\rm Pl}}\sim10^{11}~{\rm GeV}~.
\eeq
Scalar VEVs can be supersymmetric; thus, one can spontaneously break the R-symmetry at a high scale while remaining supersymmetric to the TeV scale.
Moreover, as we discuss in subsequent sections, there will also be small explicit breaking of the U$(1)_R$. It follows that the spontaneous breaking of the global R-symmetry will result in a pNGB: the R-axion.  Therefore, if QCD is the dominant source of explicit breaking, the scale $f_a$ directly determines the axion mass as standard \cite{Srednicki:1985xd} 
\beq\label{amass}
m_a\approx\frac{f_\pi m_\pi}{f_a}\approx1.2\times10^{-4}~{\rm eV}\left(\frac{10^{11}~{\rm GeV}}{f_a }\right).
\eeq
Such values of $f_a$ are typically considered desirable for cosmological considerations \cite{Patrignani:2016xqp}.

After PQ and electroweak symmetry breaking there are two neutral Goldstone bosons, one being the axion while the other is eaten by the $Z$.
Since hypercharge allows for redefinitions of the overall normalisation, if the Higgses are charged under the PQ-symmetry (cf.~DFSZ models  \cite{Zhitnitsky:1980tq,Dine:1981rt}) one must identify the physical axion. One can ensure that there is no mixing between the axion and the longitudinal mode of the $Z$ by requiring that the following constraint is satisfied  \cite{Srednicki:1985xd}
\beq\label{or}
\sum_s{Y_{s}R[s]v^2_s}=0~,
\eeq
summing over the charges and VEVs of all complex scalar fields $s$. 

This restriction is greatly simplified in the class of models we consider here since the $X$ and $Y$ fields that dominantly break the R-symmetry are gauge singlets, thus $Y_X=Y_Y=0$.
Indeed, provided that the only scalars with non-zero hypercharge that acquire VEVs are the Higgses, this requirement can be re-expressed as
\beq \label{beta}
 \frac{v^2_{u}}{v^2_{d}}= \frac{R[{H_d}]}{R[{H_u}]} = \tan^2\beta~.
\eeq
The parameterisation of our Higgs R-charges, with $R[H_u]=A\cos^2\beta$ and $R[H_d]=A\sin^2\beta$, matches the conventional use of $\tan\beta$ to be the ratio of the Higgs VEVs. 
Thus eq.~(\ref{or}) is only constraining with regards to the fact that the value of $\tan\beta$ impacts the phenomenology. In particular, perturbativity of the top Yukawa above the electroweak scale typically requires $\tan\beta \gtrsim1.2$, similarly avoiding the bottom Yukawas from blowing up constrains $\tan\beta \lesssim65$ \cite{Martin:1997ns}. The requirement $\tan\beta>1$ implies  the Higgses cannot have the same R-charge ($R[H_u]\neq R[H_d]$), unless both scalars have zero R-charge.

\section{Explict R-breaking}
\label{sec:3}

The explicit breaking of the U(1) R-symmetry by the QCD anomaly results in a relaxation of $\bar\theta$ to zero and gives the R-axion a non-zero mass, just as in the standard PQ mechanism. Thus to solve the strong CP problem one requires a U(1)${}_R\times$SU(3)${}^2$ anomaly, this is calculated via current algebra \cite{Srednicki:1985xd} by evaluating the anomaly coefficient $N=\sum_i R_i T[r_i]$, summing over the field content where $T(r_i)$ is the Dynkin index of the SU(3)${}_c$ representation $r_i$ and its associated R-charge $R_i$. For the most general model presented in Tables \ref{Tab1} \& \ref{Tab2}, the anomaly coefficient is evaluated to be
\beq
N=-\frac{3}{2}A.
\label{agg}
\eeq
Note that one adds 3 units to the U(1)${}_R\times$SU(3)${}^2$ anomaly from the gluino contribution (with $T(8)=3$), however (as we discuss in Section \ref{sec:4}) our model requires SU(3) adjoint superfields $\mb\Phi_g$ for a Dirac gluino mass and their contributions cancel against the gluino contribution. 
We observe that with the field content of Tables \ref{Tab1} \& \ref{Tab2} the R-symmetry is explicitly violated by the SU(3) anomaly, provided that $A\neq0$.

Provided that some Standard Model quark carries R-charge, one can rotate away the $\bar\theta$ angle, in the manner of DFSZ-type models \cite{Zhitnitsky:1980tq,Dine:1981rt}. From inspection of Table \ref{Tab1}, the only case in which Standard Model quarks does not carry R-charge is when $Q=1$ and $A=0$; this corresponds to the binary charge assignments which we do not consider here.

\subsection{Explicit R-breaking and Axion Potential}\label{rexplicitbreak}

Importantly, if a source other than the QCD anomaly introduces explicit breaking of the PQ symmetry then the PQ mechanism will no longer drive $\bar{\theta}$ to zero \cite{Kamionkowski:1992mf,Barr:1992qq,Holman:1992us}. This potentially spoils the PQ mechanism and reintroduces the strong CP problem. Since 
many formulations of low energy supersymmetry require explicit R-breaking, we next discuss how these considerations restrict to the 
classes of models in which one can successfully identify U(1)${}_R$ with the PQ symmetry.

The explicit breaking of the PQ symmetry due to the QCD anomaly relates the axion mass  to QCD scale and the PQ-breaking scale  via $m_a\approx f_\pi m_\pi/f_a$, cf.~eq.~(\ref{amass}).
In our case the PQ symmetry is U(1)${}_R$ and any additional explicit R-breaking leads to a corresponding new mass contribution $m_a^{\not R}$, such that
\beq
m_a= \sqrt{(m_a^{\rm QCD})^2 + (m_a^{\not R})^2}~.
\eeq
One can then parameterise the axion potential in the following manner
\cite{Barr:1992qq}
\beq
V\simeq \frac{1}{2}m^2_a a^2-m^2_a \bar{\theta} f_a a~.
\eeq
The inclusion of the second term shifts the minimum of the potential from $\langle a/f_a \rangle\equiv\bar\theta=0$ to a finite value which is parametrically \cite{Barr:1992qq,Kamionkowski:1992mf}
\beq
\bar\theta\sim\left(\frac{m_a^{\not R}{}}{m_a^{\rm QCD}{}}\right)^2< \bar\theta_{\rm exp} ~,
\eeq
where we have imposed the experimental exclusion $\bar\theta_{\rm exp}\lesssim 5.6\times10^{-11}$. This limit comes from comparing the prediction \cite{Crewther:1979pi} for the $\bar\theta$-induced neutron electric-dipole moment $d_n \simeq 5.2 \times 10^{-16} \overline{\theta} \, e \, \text{cm}$  with the experimentally observed   \cite{Baker:2006ts} upper bound $d_n<2.9\times10^{-26}~e~{\rm cm}$. 

In the models that we shall consider, in addition to the QCD anomaly, there are two potentially problematic sources of explicit R-breaking:
\begin{itemize}
\item For a global R-symmetry one anticipates Planck-suppressed higher dimension operators will violate the symmetry (unless such operators are forbidden for some reason).
\item Constant terms in the superpotential explicitly break the R-symmetry. Such terms are commonly used to tune the cosmological constant to zero.
\end{itemize}
We shall discuss the first point in Section \ref{sec:5}, specifically highlighting that the Green Schwarz mechanism provides a potential resolution. The latter issue will be explored in the remainder of this section.

Constant terms in the superpotential $W_0$ explicitly break R-symmetry \cite{Bagger:1994hh} (and earlier in \cite{Fayet:1977vd}). Such constant terms are commonly included in supergravity theories which set $\langle |W| \rangle\neq0$ to tune the cosmological constant near zero  (see e.g.~\cite{Nilles:1983ge}) via
\beq\label{cc}
\langle V \rangle = \langle |F| \rangle^2-3 \frac{\langle |W| \rangle^2}{M_{\rm Pl}^2} \exp \left[\frac{\langle K \rangle}{M_{\rm Pl}^2}\right] +\frac{1}{2}\langle |D| \rangle^2\approx 0~,
\eeq
where $F$ and $D$ are the $F$- and $D$-term scalar potentials and $K$ is the K\"ahler potential.
The typical approach to tune this sum to zero is via a constant $W_0$ in the superpotential such that $\langle |W| \rangle=W_0+\cdots$. The occurrence of $\langle |W| \rangle\neq0$ subsequently generates a gravitino mass contribution
\beq
m_{3/2}=\frac{\langle |W| \rangle}{M_{\rm Pl}^2}\exp \left[\frac{\langle K \rangle}{2M_{\rm Pl}^2}\right]~.
\eeq
By comparison to eq.~(\ref{cc}), assuming vanishing $D$-term, tuning the cosmological constant to zero, gives the familiar form $m_{3/2}\sim F/M_{\rm Pl}$. 

Explicit violation of the R-symmetry by $W_0\neq0$ gives a mass to the axion, although this is model-dependent. 
In Appendix \ref{softbreak} we find, via an explicit calculation of the supergravity scalar potential, that $m_a^{\not R}\sim  m_{3/2}$. 
It was argued by Bagger, Poppitz \& Randall \cite{Bagger:1994hh} that for generic, calculable models of dynamical supersymmetry breaking one anticipates  that $m_a^{\not R}\sim  \sqrt{m_{3/2}m_{\rm soft}}$.
Taking the optimistic view that $m_a^{\not R}\sim  m_{3/2}$ may hold in some models with $m_{3/2}\ll m_{\rm soft}$ this still leads to strong constraints on the other model parameters from the requirement that  $(m_a^{\not R}/ m_a^{\rm QCD})^2< \bar\theta_{\rm exp} $, which implies
\beq \label{maratio}
\left(\frac{m_a^{\not R}{}}{m_a^{\rm QCD}{}}\right)
& \sim  \frac{m_{3/2}  f_a}{m_\pi f_\pi}\\
&\sim  8\times10^{-6}  \left( \frac{ m_{3/2} }{10^{-6}~ {\rm eV}} \right) \left( \frac{ f_a}{10^{8}~{\rm GeV}} \right)  \\
&~\lesssim2.5\times10^{-5}~.
\eeq
PQ scales below $f_a\sim10^8$ GeV are phenomenologically problematic. 
Moreover, a gravitino mass of order $10^{-6}~ {\rm eV}$ naively implies $F=m_{3/2}M_{\rm Pl}\sim 10^3$ GeV${}^2$ in which case it seems impossible to avoid sub-TeV scale superpartners even with low scale gauge mediation.  

The conclusion is that conventional approaches towards tuning away the cosmological constant disrupt the solution to the strong CP problem involving U(1)${}_R$. The issue raised above can be viewed as a restriction on the size of $W_0$ in order to avoid generating too large of a correction to $\bar\theta$.  Our options for proceeding are either\footnote{One further alternative we do not pursue is to simply ignore the requirement that the cosmological constant is tuned to zero (since there is no mechanism to explain this tuning), in which case one can readily set $W_0=0$. One would still need to ensure that the gravitino obtained a non-zero mass.}  to~search for classes of models for which the masses induced by $W_0\neq0$ lead to $m_a^{\not R}\lesssim m_{3/2}\ll m_{\rm soft}$, or to identify R-symmetric approaches to break supersymmetry and tune away the cosmological constant with $W_0=0$. We shall discuss the latter prospect in the next section.

\subsection{Conventional Tuning of the Cosmological Constant}

To reproduce the Standard Model at low energy from supergravity one typically wishes to simultaneously break supersymmetry and tune the cosmological constant near to zero.
Before we examine an R-symmetric model due to Claudson, Hall, \& Hinchliffe \cite{Claudson:1983cr}, let us  understand how the orthodox approach accomplishes this (we follow the discussion in Nilles \cite{Nilles:1983ge}).

The structure of the $ \mathcal{N} = 1 , ~D=4 $ supergravity Lagrangian is defined by the the superpotential  $ W({\mb z_i}) $ along with the  K\"ahler potential $ K({\mb z_i}, {\mb z_i^*})$ 
(see Appendix \ref{APBBB} for a brief overview of $N =1$, $D=4$ supergravity). The tree-level scalar potential is then as a sum of $F$-term and $D$-term contributions which arise from the choice of $W$ and $K$ (see e.g.~\cite{Nilles:1983ge,Brignole:1997wnc})
\begin{equation}
V = K_{i\bar{j}} F^i\bar{F}^{\bar{j}} - 3 e^{K/M_{\rm {Pl}}^2}  \frac{|W|^2}{M_{\rm {Pl}}^2}  + \frac{1}{2}D_aD^a~,  
\label{supergravity}
\end{equation}
we shall neglect to specify $D$ for now, but the $F$-term piece is given by
\beq \label{Fterm}
F^i &= e^{K/2M_{\rm {Pl}}^2} K^{i\bar{j}}\left(  \partial_{\bar{j}} \bar{W} +   \partial_{\bar{j}} K \frac{\bar{W}}{M_{\rm {Pl}}^2}\right)~,\\
\bar{F}^{\bar{j}} &= e^{K/2M_{\rm {Pl}}^2} K^{i\bar{j}}\left(  \partial_i W +   \partial_i K \frac{W}{M_{\rm {Pl}}^2}\right)~,
\eeq
where $K_{i \bar{j}} = \frac{\partial^2 K}{\partial \phi^i \partial \bar{\phi}^{\bar{j}}}$ is the K\"ahler metric. Supersymmetry is broken if either $\langle F_i \rangle\neq0$ or $\langle D_a \rangle\neq0$. The requirement that the cosmological constant is tuned near zero implies $\langle V \rangle\approx0$.

To see how this works in conventional approaches we employ the prototypical hidden sector supersymmetry breaking model with the Polonyi superpotential \cite{Polonyi:1977pj} and minimal K\"ahler potential
\beq
W_P(z)&=m^2({\mb z}+W_0)~, \\ 
K&={\mb z}{\mb z^*}~.
\eeq
This superpotential involves a single chiral superfield $z$ and constant dimensionful parameters  $m$ and $W_0$. This K\"ahler potential implies $K_{zz^*}=K^{zz^*}=1$. Evaluating the $F$-term yields
\beq
\overline{F}^{z^*} = e^{\frac{|z|^2}{M_{\rm {Pl}}^2}} \frac{m^2}{M_{\rm Pl}^2} \left( M_{\rm Pl}^2+ z^* (z + W_0)\right)~.
\eeq
In the case that there is no $D$-term breaking the scalar potential is given by 
\beq
V = 
 \frac{m^4}{M_{\rm {Pl}}^4} e^{\frac{|z|^2}{M_{\rm {Pl}}^2}} \left[  | M_{\rm Pl}^2 +  z^* (z + W_0)|^2
-3M_{\rm Pl}^2|z+W_0|^2)
 \right]~.
\eeq
Following Nilles \cite{Nilles:1983ge}, let us look for the parameters needed in Polonyi superpotential $W_P$ for the scalar potential to have a minimum at zero $\langle V'(z_0)\rangle =0$ with $V(z_0)=0$, as well as $F(z_0)\neq0$ to ensure supersymmetry is broken.
We start by computing the derivative of $ V $ with respect to $ z $.
Requiring that  $\frac{\partial V}{\partial z}=0$ implies
\beq\label{APC1}
&\frac{z^*}{M^2_{\text{Pl}}} \left( |\Theta|^2 - 3M^2_{\text{Pl}} |z + W_0|^2 \right)  \\
&\quad + \Theta(z^* + W_0)+ z^* \Theta^* - 3M^2_{\text{Pl}} (z^* + W_0) = 0~,
\eeq
where $\Theta\equiv M^2_{\text{Pl}} + z^* (z + W_0)$.
This expression gives the critical points of the potential,
which are found for 
\beq\label{APC2}
|\Theta|
= \sqrt{3} M_{\rm Pl} |z + W_0|.
\eeq
We can substitute $\Theta$ into eq.~(\ref{APC2}), then to simultaneously satisfy eq.~(\ref{APC1}) implying $ V' = 0 $ and eq.~(\ref{APC2}) implying $ V = 0 $ leads to the requirement 
\beq
(W_0 + z + z^*) \left[ M_{\text{Pl}}^2 + z^* (z + W_0) \right] = 3M_{\text{Pl}}^2 (z^* + W_0).
\eeq
For $|W_0|=(2-\sqrt{3})M_{\text{Pl}}$ the potential has an absolute minimum at $V=0$ and the VEVs of $z$ and $F$ are found to be \cite{Nilles:1983ge}
\beq
\langle z \rangle\equiv z_0 &= (\sqrt{3}-1)M_{\text{Pl}} ~,\\
F(z_0)&=\sqrt{3} m^2 ~, \\
 V(z_0)&=0~.
\eeq
Thus supersymmetry is broken. Notably, the superpotential $W_P$ is unnatural since it does not contain ${\mb z}^2$ and ${\mb z}^3$ terms.  The form of $W_P$ can only be made natural via an R-symmetry with $R[{\mb z}]=2$, however, this R-symmetry would be explicitly violated for $W_0\neq0$. 
We also highlight that certain supergravity completions require gauged R-symmetries and these seem inconsistent with the constant in the superpotential. 
For these reasons it is attractive to identify a manner of tuning the cosmological constant to zero without explicit R-violations in either the K\"ahler potential or superpotential.

\subsection{R-symmetric Tuning of the Cosmological Constant}
\label{AC3}

The above picture underlies most phenomenological approaches to supergravity, since it cleanly tunes away the cosmological constant at the cost of breaking U(1)${}_R$ (a symmetry which is also phenomenologically bothersome). However, one can look to tune the cosmological constant for $W_0=0$, such that there are no explicit tree level violations of the R-symmetry, while breaking supersymmetry and generating a mass for the gravitino. Claudson, Hall, \& Hinchliffe  \cite{Claudson:1983cr} gave the first example of such a scenario by using the parameters of a non-minimal K\"ahler potential to tune the cosmological constant to zero. For variations on this class of R-symmetric models see \cite{Pallis:2018xmt,Pallis:2020dxz,Aldabergenov:2019ngl,Kitazawa:1999su,Kitazawa:2000cz,Kitazawa:2000ft}, and also the related work of \cite{Antoniadis:2014iea,Antoniadis:2014hfa,Antoniadis:2015mna}.

A requirement of this model is that the  superpotential should be separated into visible sector and hidden sector, both of which are R-symmetric, with have no renormalizable cross terms. The separation in the superpotential between visible and hidden sector components is naturally enforced through a combination of the continuous R-symmetry and gauge symmetries. 
Specifically, Claudson, Hall, \& Hinchliffe \cite{Claudson:1983cr} considered the following R-symmetric superpotential and (non-minimal) K\"ahler potential 
\beq \label{Hall}
W_R(z)=m^2 {\mb z}~, \qquad K_R=  {\mb z}\bar{{\mb z}} + \frac{a}{2} \frac{({\mb z}\bar{{\mb z}} )^2}{M_{\text{P}}^2} ~.
\eeq
Assuming no $D$-term contribution the scalar potential is given by
\begin{equation} 
 V  = e^{\left(\frac{z \bar{z}}{M_{\text{Pl}}^2} + \frac{a}{2} \frac{(z \bar{z})^2}{M_{\text{Pl}}^4}\right)} m^4 \left[ \frac{\left( 1 + \frac{z \bar{z}}{M_{\text{Pl}}^2} + \frac{a z^2 \bar{z}^2}{M_{\text{Pl}}^4} \right)^2}{ 1 + \frac{2a}{M_{\text{Pl}}^2} z \bar{z}} 
  - \frac{3  |z|^2}{M_{\text{Pl}}^2} \right] ~.
  \label{VAC3}
\end{equation} 
We can simplify the potential by writing $r=|z|^2/M_{\text{Pl}}^2$
 \beq
  V = e^{r+ \frac{a}{2} r^2}m^4\left[ \frac{1}{ 1 + 2ar}   (1 + r + a r^2)^2 - 3 r   \right] ~.
  \label{VVVVV}
 \eeq
First note that the condition that $V=0$ can be expressed as
\beq
a=\frac{2-r\pm \sqrt{3(1-r)}}{r^2}~.
\label{aaaaa}
 \eeq
Observe that $r\leq1$ for real $a$ which also implies $a$ is positive.
The minimum of the potential may be tuned to zero by choosing $a$ carefully, and  \cite{Claudson:1983cr}  derives an exact form for this special value $a=a_*$ given by
\beq
a_* &= \frac{1}{4} + \frac{3}{16} \left[ (2 + \sqrt{3})^{1/3} + (2 - \sqrt{3})^{1/3} \right]\\
&\approx0.6617168772710588411561...
\eeq
This particular value of $a$ implies the  VEV of $z$ is given by $\langle z\rangle^2=r_*M_{\text{Pl}}^2$ with
\beq\label{r*}
r_*&=
 \frac{1}{\frac{1}{2}+\frac{1}{4} \left[(7-4 \sqrt{3})^{1/3}+(7+4 \sqrt{3})^{1/3}\right]}\\
&\approx 0.8295932221425617511993...
\eeq
We have numerically verified that the potential of eq.~(\ref{VVVVV}) with $a=a_*$, call this $V_{*}$, leads to  $\langle z\rangle^2\equiv z_0=r_*M_{\rm Pl}^2$ and $V_{*}$ is found to be positive definite. Curtailing $a_*$ and $r_*$ at the shown decimal place and evaluating $V_{*}$ we find $V_*(z_0)\approx0\times m^4$ up to numerical precision. This numerical prefactor can be tuned arbitrarily small by taking a longer decimal expansion. One potential concern,  raised in \cite{Nilles:1983ge}, is that since $r\not\ll1$ one may worry about corrections beyond the tree-level scalar potential; considerations of this are beyond the scope of this work.

From the $z$ VEV, one can see that supersymmetry is broken by calculating the $F$-term 
\beq
| F_z(z_0)  |\label{Fterm0}
&=  \exp\left[\frac{2r_*+ar_*}{4}\right] \left( 1 + 2a_* r_* \right)^{-1}  \left( 1 + r_* + a_* r_*^2 \right) m^2\\
& \approx 1.8m^2~,
\eeq
and the gravitino mass in this model is given by
\beq\label{HallG}
 m_{3/2} &= \left\langle e^{K/2M_{\text{Pl}}^2} \frac{W}{M_{\text{Pl}}^2} \right\rangle\\
& =\exp\left[\frac{2r_*+ar_*}{4}\right]\frac{m^2 \sqrt{r_*}}{M_{\text{Pl}}}\approx1.6\frac{m^2 }{M_{\text{Pl}}}~.
\eeq
Given the $F$-term of eq.~(\ref{Fterm0}) the normalised soft scalar masses are generated by gravity mediation of order $m_0\sim m_{3/2}$ as can be seen via \cite{Brignole:1997wnc} 
\beq 
 m_0^2 
 &= m_{3/2}^2 - F^i F^{\overline{j}} \partial_i \partial_{\overline{j}} \log \tilde{K}~,
  \eeq
 where $\tilde{K}$ appears in the K\"ahler potential as follows $ K\supset \tilde{K}_{\overline{i}j}\overline{\Phi^i}\Phi^j$  (see  eq.~(\ref{W}) in Appendix \ref{softbreak}). As explained in \cite{Claudson:1983cr} the same method employed above to tune the cosmological constant can be used when MSSM fields are added to the theory with a minor modification. If one assumes that the K\"ahler potential is a function only of $\sigma \equiv {\mb z}\bar{{\mb z}} + \overline{\Phi}^{i}\Phi_i$ where $\Phi_i$ denotes a MSSM superfield any by taking the K\"ahler potential to be that of eq.~(\ref{Hall}) with ${\mb z}\bar{{\mb z}}$ replaced by  $\sigma$ one can cancel the cosmological constant by following through the same steps as above. From this form of the K\"ahler potential one has $\tilde{K}=1+ \frac{a({\mb z}\bar{{\mb z}})^2}{M^2_{\rm Pl}} $ giving

\beq
m^2_{0}=m^2_{3/2}\left(1-\left[\frac{a_*(1+2a_* r_*)^{-2}(1+r_*+a_*r^2_*)}{r_*(1+a_*r_*)^2}\right]\right)~.
\eeq
Evaluating numerically: $m_0\approx 0.78 m_{3/2}\approx 1.25 m^2/M_{\rm Pl}$ where $m$ is the scale of the hidden sector superpotential, which can be freely chosen. For TeV sfermions we require $m\sim10^{11}$ GeV. It is curious that this scale coincides with the energy scale commonly favoured for the axion $f_a$, but at present, this is accidental.

In the absence of explicit R-breaking the Majorana gaugino masses and A/B-terms will be absent prior to spontaneous R-breaking (we discuss this further in Section~\ref{sec:4}). 
Observe that both the superpotential and K\"ahler potential are R-symmetric, unlike the Polonyi example with $W_P=m^2({\mb z}+W_0)$, thus this model provides \textit{proof of principle} that one does not lose the capacity to tune-away the cosmological constant or give mass to the gravitino while avoiding tree level explicit R-violation.   
While the VEV of $z$ spontaneously breaks the R-symmetry in the hidden sector, there is no explicit R-breaking, and the spontaneous R-breaking is only communicated to the visible sector gravitationally.  
To avoid the transmission of Planck scale spontaneous R-breaking due to $\langle z\rangle\neq0$ to the visible sector it is important that the $z$ field is highly decoupled, as we discuss next.

\subsection{Dominant Visible Sector R-breaking}
\label{se3.4}
In the DSFZ axion  \cite{Zhitnitsky:1980tq,Dine:1981rt} the field $S$ that sources the dominant PQ breaking is connected to the quark-gluons triangle diagram via the operator $SH_uH_d$. In our model the $z$ fields do not couple to quark-gluons triangle diagram by any renormalisable operators, however, we should check if there may exist non-renormalisable operators involving fields that acquire VEVs.
Since the hidden sector is only coupled to the visible sector gravitationally (due to the R-symmetric structure of our model of Section \ref{AC3}) one might na\"ively anticipate that the hidden sector global U(1) involving ${\mb z}$ and the visible sector global U(1) are distinct. For any other global U(1) this would be true, but this is not the case for the R-symmetry since it rotates the superspace co-ordinates $\vartheta$ in both sectors.

From eq.~(\ref{supergravity}) one can derive general expressions for terms that generate the soft scalar masses after supersymmetry breaking assuming the hidden sector K\"ahler and superpotential of eq.~(\ref{Hall}). If the scalar Lagrangian from supergravity contains terms of the form 
 \beq
 \mathcal{L}\supset \mathcal{Z}(z,\overline{z})H_{u}H_{d}~,
 \eeq
where $\mathcal{Z}$ is a polynomial function with $ \mathcal{Z}(z,\overline{z})\neq \mathcal{Z}(|z|^2)$, these terms can potentially drag up the axion decay constant to $f_{a} \sim \langle z \rangle \sim M_{\rm Pl}$. 
 Such terms can arise from either the superpotential or the K\"ahler potential. Moreover, upon UV completion into supergravity they correspond to B-terms in the general supergravity scalar potential (see~Appendix \ref{softbreak}). 

 Let us first consider higher dimension terms in the superpotential of the form
\beq
 W \supset \frac{{\mb z}^p}{M_{\rm Pl}^{r}}\mathcal{O}({\mb H_u},{\mb H_d},{\mb X},{\mb Y})~,
\eeq
 where $\mathcal{O}({\mb H_u},{\mb H_d},{\mb X},{\mb Y})$ is some holomorphic operator suppressed by $M_{\rm Pl}$ to some appropriate power $r$. 
Specifically, consider some visible sector operator  $\mathcal{O}_{nm}={\mb X}^{n}{\mb Y}^{m}{\mb H}_{u}{\mb H}_{d}$.
Potentially problematic R-breaking operators can arise unless $R[{\mb z}^p\mathcal{O}_{nm}]\neq2$ for all positive integer values of $p$. 
The reason that this condition must be satisfied for all $p\in \mathbb{N}$ is that there is no suppression  for larger values of $p$  since  $(\langle z\rangle/M_{\rm Pl})^{p}\sim1$ for $p\lesssim20$. 
Using the R-charge assignments of eq.~(\ref{eq7}) we see that the operator $\mathcal{O}_{nm}$ is forbidden provided 
\beq\label{pcond}
2p \neq A\left(\frac{3m}{2}-\frac{n}{2}-1\right) -2m+2~.
\eeq
With $A\not\in \mathbb{Z}$ there will generally be no $p\in\mathbb{N}$ that satisfies this condition,  hence avoiding these problematic operators.
Moreover, we require that $m=n=0$ terms are strictly forbidden as ${\mb z}^p{\mb H_u}{\mb H_d}$ in the superpotential disrupt the framework of Section \ref{AC3}. This separation of hidden and visible sector fields can be assured provided that $A\neq0,-2,-4,\cdots$

Conversely, suppose that the value of $A$ is such that a potentially problematic operators arises from the superpotential of the form ${\mb z}^p{\mb X}^{n}{\mb Y}^{m}{\mb H}_{u}{\mb H}_{d}$ it follows that the scalar potential after VEV insertions, contains soft supersymmetry breaking DFSZ-like terms of the form
\beq
\mathcal{L}\supset
\frac{m_{3/2}}{M_{\rm Pl}}\left(\frac{m_{\rm soft}\Lambda}{M_{\rm Pl}^2}\right)^{\frac{m+n}{2}}\left(\frac{z}{M_{\rm Pl}}\right)^{p-2}z^2H_uH_d ~.\eeq
If we consider the relative size of the operators ${\mb X}{\mb Y}{\mb H_u}{\mb H_d}$ and ${\mb z}^p{\mb X}^{n}{\mb Y}^{m}{\mb H}_{u}{\mb H}_{d}$ we estimate their contributions to the axion decay constant are given by
\beq
f_a=f_{\rm vis}\sqrt{1+\Delta_W}~,
\eeq
writing $f_{\rm vis}\sim\sqrt{\langle X\rangle^2+\langle Y\rangle^2}$, and using the relation $\langle X\rangle,\langle Y\rangle\sim\sqrt{m_{\rm soft}\Lambda}$ from eq.~(\ref{<X>}), we also define
\beq
 \Delta_W=\frac{m^{2}_{3/2}}{M^2_{\rm Pl}}\frac{\langle z\rangle^2}{f_{\rm vis}^2}\left(\frac{m_{\rm soft}\Lambda}{M_{\rm Pl}^2}\right)^{m+n}\left(\frac{\Lambda}{m_{\rm soft}}\right)^2\left(\frac{\langle z\rangle}{M_{\rm Pl}}\right)^{2(p-2)}~.
\eeq
The $\langle z\rangle$ factor is typical $\mathcal{O}(1)$ since $(\langle z\rangle/M_{\rm Pl})^2\equiv r_*\approx0.8$ from eq.~(\ref{r*}).
If we take $\Lambda=M_{\rm Pl}$ and $m_{\rm soft}=m_{3/2}$ and neglect the $\langle z\rangle$ factor we obtain
\beq
\Delta_W
\sim\left(\frac{M_{\rm Pl}}{f_{\rm vis}}\right)^2\left(\frac{m_{\rm soft}}{M_{\rm Pl}}\right)^{m+n}~.
\eeq 
For operators with $m+n>1$ we find that $\Delta_W$ is negligible, thus we only need to choose $A$ such that the operators $\mathcal{O}_{nm}$ with $m+n\leq1$ are forbidden. In particular, for $m=n=0$ one finds $\Delta_W=(M_{\rm Pl}/f_{\rm vis})^2$, thus $f_a=M_{\rm Pl}$, this is because the Lagrangian has operators of the form $z^p H_uH_d$ which break the visible sector U(1)${}_R$ at the Planck scale. 

We should also consider operators from the K\"ahler potential,\footnote{Applying the general supergravity soft terms (see Appendix \ref{softbreak}) to the model of Section \ref{AC3} we find that most contributions depend only on the magnitude of the hidden-sector field, $|z|$, and not the phase. Hence most of these potentially worrisome operators will not contribute to the PQ mechanism. This is largely due to the R-symmetry and the simplicity of the hidden-sector superpotential and K\"ahler potential (cf.~eq.~(\ref{Hall})).} these have the form
 \beq \label{Kssb}
 K \supset \frac{{\mb z}^p({\mb z}^*)^q}{M_{\rm Pl}^{r}}\mathcal{O}({\mb H_u},{\mb H_d},{\mb X},{\mb Y})~.
 \eeq
The K\"ahler potential appears in the Lagrangian as $\int d^4 \theta ~K$, unlike the superpotential its measure $d^4 \theta$ is an R-invariant, thus K\"ahler terms must have zero net R-charge to be R-symmetric. 
Suppose that a visible sector operator has R-charge  $R[\mathcal{O}({\mb H_u},{\mb H_d},{\mb X},{\mb Y})]=0$ then R invariance of the K\"ahler potential requires $p=q$ in eq.~(\ref{Kssb}). In this case the scalar terms arising from the K\"ahler potential linking the two sectors will only involve the hidden sector scalar $z$ as a function
\beq
z^p(z^*)^q\big|_{p=q}=|{z}|^{2p}=\rho_z^{2p}~,
\eeq
where we write $z=\rho_z e^{ia_z}$. Since this is not a function of $a_z$ it will not lead to an $a_z G^{\mu\nu}\widetilde{G}_{\mu\nu}$ coupling and thus will not lift $f_a$ to the $\langle z \rangle\sim M_{\rm Pl}$.
In contrast, for operators with $q +\delta = p$ one has
\beq
z^p(z^*)^q\big|_{q +\delta = p}=|{z}|^{2p}(z^*)^{-\delta} =\rho_z^{2p-\delta}\exp(i\delta a_z).
\eeq
Thus operators of the form of eq.~(\ref{Kssb}) with $p\neq q$ (i.e. $k\neq0$) do lead to an $a_z G^{\mu\nu}\widetilde{G}_{\mu\nu}$ coupling and might potentially lead to a $z$-induced correction of $f_a$, although as we show below the impact of these K\"ahler operators is typically negligible.

Specifically, consider the operator from the K\"ahler potential ${\mb z}^p({\mb z}^*)^q{\mb X}^{n}{\mb Y}^{m}{\mb H}_{u}{\mb H}_{d}$, from eq.~(\ref{eq7}) we see that such the operators are R-invariant if 
\beq\label{pqcond}
\delta \equiv p-q = A\left(\frac{3m}{2}-\frac{n}{2}-1\right) -2m~.
\eeq
Similar to the superpotential condition of eq.~(\ref{pcond}), with $A\not\in \mathbb{Z}$ there will generally be no $\delta\in\mathbb{N}$ that satisfies this charge constraint, and thus these operators are forbidden.

Suppose that the R-charges are such that the leading K\"ahler potential term that connects the visible and hidden sector is of the form $({\mb z^*})^q{\mb X}^{n}{\mb Y}^{m}{\mb H}_{u}{\mb H}_{d}$ (this form give the strongest correction to $f_a$).  An example is $\mb{z^*H_uH_d XY}$ which can always be formed by appending $z*$ to the $R=2$ terms of $W_{\mu}$ given in eqn.~(\ref{Hitoshi}) and append ${\mb z^*}$ to form an R-invariant K\"ahler operator.

The DSFZ-like term can calculated from the supergravity soft breaking $B$-term given in eq.~(\ref{Bterm}) of Appendix \ref{softbreak} leading to 
\beq
\mathcal{L}\supset
\frac{m^{2}_{3/2}}{M^2_{\rm Pl}} &
\left(\frac{m_{\rm soft}\Lambda}{M_{\rm Pl}^2}\right)^{\frac{m+n}{2}}\left(\frac{z}{M_{\rm Pl}}\right)^{\delta-2}z^2H_uH_d\\[5pt]
& \Rightarrow\quad f_a=f_{\rm vis}\sqrt{1+\Delta_K}~,
\eeq
where
\beq\hspace*{-2mm}
 \Delta_K=\frac{\langle z^2\rangle}{f_{\rm PQ}^2}\left(\frac{m_{3/2}}{M_{\rm Pl}}\right)^4\left(\frac{m_{\rm soft}\Lambda}{M_{\rm Pl}^2}\right)^{m+n}\left(\frac{\Lambda}{m_{\rm soft}}\right)^2\left(\frac{\langle z\rangle}{M_{\rm Pl}}\right)^{2(\delta-2)}~.
\footnotesize\eeq
Notably, making the natural assumptions $\Lambda=M_{\rm Pl}$ and $m_{\rm soft}=m_{3/2}$ 
 even in the worst case, with $n=m=0$, this correction is $\Delta_K\sim (m_{\rm soft}/M_{\rm Pl})\sim 10^{-15}$. Thus K\"ahler operators are typically not a concern with regards to dominating the visible sector R-breaking.

However, the case $n=m=0$ corresponds to a new term in the superpotential of the form ${\mb z}^p{\mb z^*}^q{\mb H}_u{\mb H}_d$ which complicates (and may disrupt) the framework of Section \ref{AC3}. Since it unclear if such terms in the K\"ahler potential can be accommodated while tuning the potential to zero, we will require $R[{\mb z}^p{\mb z^*}^q{\mb H}_u{\mb H}_d]\neq0$ for all positive integer values of $p$ or $q$. This implies the restriction $|A|\not\in \mathbb{Z}^{\rm even}$.

In summary, to avoid $z$ breaking dragging $f_a$ to the Planck scale all one need is a choice of $A$ such that $p$ satisfies eq.~(\ref{pcond}) for $m+n\leq1$, and $|A|\not\in\mathbb{Z}^{\rm even}$ to avoid disrupting the separation of hidden and visible sectors required in Section \ref{AC3}.
If these two conditions are satisfied, the dominant source of spontaneous R-breaking in the visible sector will come from the $X$, and $Y$ scalar VEVs (according to eq.~(\ref{Hitoshi})) and the sole source of explicit R-breaking can potentially come from the QCD anomaly.\footnote{One typically expects Planck-suppressed operators to violate all continuous global symmetries \cite{Banks:2010zn}, including U(1)${}_R$, however, if the R-symmetry is gauged in a UV completion to supergravity then these R-violating Planck-suppressed operators will be absent (we discuss this in detail in Section \ref{sec:5}).}

\subsection{$F$-term and $D$-term Breaking with an R-symmetry}
\label{3.5}

One can generalise the model of  \cite{Claudson:1983cr} to include $D$-term supersymmetry breaking. Since in Section \ref{sec:5} our ambition is to gauge\footnote{Local R-symmetries can only be realised within supergravity and string theory. For global supersymmetry, R-symmetries are also required to be to global transformations (see e.g.~\cite{Chamseddine:1995gb}).} U(1)${}_R$ in the UV such $D$-term potentials can arise. As $D$-terms are not inherently R-violating this is attractive.  
Moreover, it is known that constant Fayet-Iliopoulos (FI) terms that contribute to $D$-terms in supergravity must generally be associated to gauged R-symmetry  \cite{Barbieri:1982ac,Freedman:1976uk,Stelle:1978wj} (see \cite{Cribiori:2017laj} for caveats). 
Indeed, scenarios in which supersymmetry breaking  arises through a combination of hidden sector $F$-terms and $D$-terms appear in the literature, for instance, \cite{Itoyama:2011zi,Dienes:2008gj,Dumitrescu:2010ca,Azeyanagi:2011uc,Nakano:1994sw,Antoniadis:2015mna,Dudas:2005vv,Antoniadis:2014hfa,Antoniadis:2014iea}.

Thus let us briefly discuss the prospect of adding $D$-term supersymmetry breaking to the framework of  \cite{Claudson:1983cr}. As noted in eq.~(\ref{supergravity}) the $D$-term contribution associated with a U(1) gauge symmetry  to the scalar potential  is 
\beq
 V_D = \frac{1}{2}g^2 D^2 ~
\eeq
where $g$ is the gauge coupling.
For a U$(1)_R$ gauge symmetry $W \rightarrow e^{i\xi \Lambda(x)}W$, the $D$-term is given by (see Appendix.~\ref{scalarpot})
\beq
\label{DDD2}
 D = -\sum_i R_i {\mb \Psi}_i \frac{\partial K}{\partial {\mb \Psi_i}} + \xi M_{\rm Pl}^2~,
\eeq 
where  $R_i$  are the charges of the fields  ${\mb \Psi}$ which carry U(1) symmetry and $\xi$  is the FI term.  
Supposing that U(1)${}_R$ is gauged in the UV, we shall re-examine the model of Section~\ref{AC3} involving the hidden sector superfield ${\mb z}$ with $R[{\mb z}]=\xi$ (this is a toy model as we are neglecting visible sector superfields with R-charge). The $D$-term contribution from $z$ alone is 
\beq\label{DDD}
\frac{1}{2}g^2 D^2 
&= \frac{1}{2}g^2 \left( \xi |z|^2 \left( 1 + \frac{a |z|^2}{M_{\rm Pl}^2} \right) + \xi M_{\rm Pl}^2\right)^2 \\
&=g^2\frac{M_{\rm Pl}^4\xi^2}{2}\left[
r^2(1 +2a r+a^2r^2) +2 \left( r + a r^2 \right)+ 1 \right]~.
\eeq
Thus the full scalar potential is the sum of eq.~(\ref{VAC3})  and eq.~(\ref{DDD}).
We should look to choose $a$ such that $V(z_0)=V_D(z_0)+V_F(z_0)\approx0$. Solving this equation is complicated, especially since  $ |V_D| \propto |z|^4 \propto  M_{\rm Pl}^4$ while from inspection of eq.~(\ref{VVVVV}) we anticipate $ |V_F| \propto m^4$. 

One can, however, tune both $a$ and $\xi$ such that both $V_F$ and $V_D$ are individually small and $V=V_F+V_D\approx 0$.  Specifically, if we take $a=a_*(1+\epsilon)$ for $|\epsilon|\ll1$, this implies $V_F\approx0 -m^4\epsilon +\mathcal{O}(\epsilon^2)$. We should tune $\xi$ not so that the $D$-term potential vanishes but such that it gives an equally small (opposing) contribution. Specifically, such that $D\sim  m^2\sqrt{2\epsilon}$, thus $V_D\sim m^4\epsilon$. 
Notice $V_D$ always give a positive contribution thus we require $\epsilon>0$ and since $\xi$ appears as a prefactor in eq.~(\ref{DDD}) to arrange for $V_F+V_D\approx0$ it is required that $\xi$ is $\xi\simeq \sqrt{2\epsilon} (m/M_{\rm Pl})^2$. We will discuss in Section \ref{sec:cancel} the freedom and restrictions for choosing $\xi$. Moreover, in a full model there will be additional U(1) D-terms with contributions from all of U(1) charged field content which will introduce further freedom.

In tuning $\xi$ the cosmological constant is cancelled to an acceptable degree and the $D$ and $F$ terms are comparable with $F\sim D\sim m^2$.
Note that there are existing examples in the literature in which $V_F$ and $V_D$ cancel against each other, see for instance \cite{Nakano:1994sw,Mohapatra:1996in,Chamseddine:1995gb,Antoniadis:2015mna,Antoniadis:2014iea,Antoniadis:2014hfa,Dudas:2005vv}. 
Such a tuning of $\xi$ may seem unaesthetic from a low energy perspective since $\xi$ sets many of the R charges of the theory (see Appendix \ref{superconformal}). In certain cases one can choose the FI term such that $V_D$ vanishes, leaving only $F$-term supersymmetry breaking, which would recover the set-up of Section~\ref{AC3}. This will be the case in our discussion is Section 5, where we discuss the gauging of the R-symmetry via the GS mechanism. This leads to extra contributions to the $D$-term not proportional to $\xi$ allowing for the desired tuning between the $F$ and $D$-terms without tuning $\xi$, see Section \ref{axdecay} for more complete discussion. 

One could also consider pure $D$-term breaking. Scenarios in which the cosmological constant is tuned to zero with $F=0$ have been constructed \cite{Dvali:1997sf,Villadoro:2005yq,Dudas:2005vv}, however, all of these involve explicit R-breaking in the superpotential. An R-symmetric model could have $W=m^2{\mb z}$ but one must then identify a K\"ahler potential such that $F=0$, but $D\neq0$, and $m_{\rm 3/2}\propto \langle W\rangle\neq 0$.  It remains unclear whether this can be realised, thus we will not discuss this further.


\section{Model Building}
\label{sec:4}

We next explore phenomenological issues such as giving gauginos phenomenologically acceptable masses, and the prospect of dangerous proton decay operators.

\subsection{Gaugino Mass Terms}
\label{3.1}

Continuous R-symmetries famously forbid $A$-terms, the $\mu$ term, and gaugino masses \cite{Hall:1990hq}. In the case that U(1)${}_R$ is (almost) exact this presents a problem for arranging for phenomenologically viable gaugino masses. As we show below although Majorana masses are generated after spontaneous R-breaking at an intermediate scale (via the VEVs of $X$ and $Y$) these still exhibit problematic suppression. Specifically, the size of the Majorana gaugino masses arise through the interplay of an R-breaking scalar VEV $\langle X \rangle\sim f_a\sim\sqrt{m_{\rm soft}\Lambda}$ (from eq.~(\ref{<X>})) and a supersymmetry breaking $F$-term $F$ 
\beq\label{WW}
m_{1/2}\sim \frac{f_a F}{M \Lambda} 
 \sim \frac{F}{M} \sqrt{\frac{F}{M\Lambda}}~,
\eeq
 where $M$ is the mass scale at which supersymmetry breaking is communicated to the visible sector. Within the framework of Section \ref{sec:3} we have assumed gravity mediated supersymmetry breaking for which $M=M_{\rm Pl}$. Generalisations may be possible with other mediation scenarios, in which case $M$ could be some intermediate scale $10^5~{\rm GeV}\lesssim M <M_{\rm Pl}$.

For gauge mediation, in order for this mass term to arise without further suppression one requires that the messengers have appropriate R-charge assignments (compare with \cite{Fu:2023sfk}). In the case of gravity mediation the form of eq.~(\ref{WW}) requires a specific form of the gauge kinetic function $\mathcal{F}_{ab}=\delta_{ab}\mathcal{F}_{a}$. 
We can better understand the dimensional analysis that leads to eq.~(\ref{WW}) by appealing to supergravity. In gravity mediation the (normalised) gaugino masses are generated after supersymmetry breaking by
\beq
M_{a} = \frac{1}{2}({\rm Re} \mathcal{F}_{a})^{-1} F^{m}\partial_{m}\mathcal{F}_a~.
\eeq
To generate gaugino masses  $\mathcal{F}_{a}= g^{-2}(1+\mathcal{G}/M_{\rm Pl}^n)$ where $\mathcal{G}$ is an R-invariant holomorphic Planck suppressed operator. The common choice of $\mathcal{G}= {\mb z}$ is unsuitable for the R-symmetric hidden sector model of Section \ref{AC3} as $R[{\mb z}]=2$. Thus instead the gauge kinetic function must be a less trivial function\footnote{As discussed in Section \ref{gs_sec} if U(1)${}_R$ is gauged via the Green-Schwarz mechanism the leading contribution to the gauge kinetic function is proportional Green-Schwarz modulus.} of the R-charged SM-singlet scalars $\mathcal{G}=\mathcal{G}({\mb X},{\mb Y},{\mb z})$. For example, one might choose $\mathcal{G}={\mb z}{\mb Y}$ and $R[{\mb Y}]=-2$ (corresponding to $A=8/3$) which leads to $R[\mathcal{G}]= 0$. Such a choice of $\mathcal{G}$ generate gaugino masses of the scale given in eq.~(\ref{WW}) after spontaneous breaking of both supersymmetry and R-symmetry (taking $z \rightarrow F_{z}$, and $ Y \rightarrow {f_{a}}$). Thus, in this gravity mediation scenario,  due to R-invariance gaugino masses are suppressed by a factor of $f_a/M_{\rm Pl}$ compared to the soft scalar masses.

  Furthermore, for the effective field theory in eq.~(\ref{Hitoshi}) to be consistent we should require $f_a\lesssim \Lambda$. Since we expect that $f_a\gg m_Z$ (for instance $f_a\sim10^{10}$ GeV) one finds that eq.~(\ref{WW}) does not lead to sufficiently heavy gaugino masses to be compatible with collider searches
\beq
m_{1/2}  \sim 10~{\rm GeV} \left(\frac{10^{10}~{\rm GeV}}{\Lambda}\right)^{1/2} \left(\frac{m_{\rm soft}}{10^4~{\rm GeV}}\right)^{3/2}~.
\eeq
However, once the R-symmetry is spontaneously broken in the visible sector Majorana gaugino mass terms can be radiatively generated  \cite{Farrar:1982te,Fayet:1978qc}. Specifically, there is an additional mass contribution for the electroweakinos due to Higgsino loops with $\mu$-term insertions  
\beq\label{eee}
m_{\rm EWino}\sim\frac{\mu}{16\pi^2}\sim \frac{m_{\rm soft}}{16\pi^2}~.
\eeq
This is still rather light, the suppression factor being $\mathcal{O}(100)$ compared to the soft breaking scale. The situation is markedly worse for the gluinos $\widetilde{g}$ which arise with an extra suppression of $m_{\rm soft}/\Lambda$, since they require a squark-quark-Higgs coupling \cite{Banks:1993en,Farrar:1994ce}. This causes the gluino mass contribution to be highly suppressed 
\beq
\label{glue}
m_{\widetilde{g}}\sim \left(\frac{m_{\rm soft}}{16\pi^2 \Lambda}\right) m_{\rm soft} \ll m_{\rm soft}~.
\eeq

We see three possible routes to obtaining acceptable gaugino masses:
\begin{itemize}
\item[(a).] Ignoring the hierarchy problem we can take $m_{\rm soft}\gg 10$ TeV, leading to split or high scale supersymmetry scenarios \cite{Giudice:2004tc,Arkani-Hamed:2004ymt,Wells:2004di,Hall:2009nd} (for an R-symmetric implementation see \cite{Unwin:2012fj}).
\item[(b).] We could abandon the Kim-Nilles mechanism.
\item[(c).] New physics can lead to comparable gaugino and sfermion masses.
\end{itemize}

TeV scale supersymmetry remains highly attractive for resolving the hierarchy problem, thus let us skip (a) and focus on the other possibilities.
To understand point (b) consider a different calculation in which we forget the Kim-Nilles model which connected $\langle X \rangle$ to $m_{\rm soft}$ and allow $\epsilon:=\langle X \rangle/\Lambda$ to be a free parameter, then one has $m_{1/2}\sim \epsilon m_{\rm soft}$. For instance, one could generate the $\mu$ term R-symmetrically as in \cite{Kribs:2007ac}. However, one would need to start over with the model-building and would lose this appealing structure that links $\mu$, $m_{\rm soft}$, and $f_a$.
Therefore, since we wish to retain both low-scale supersymmetry and the Kim-Nilles mechanism, we will pursue (c) and look for a new source for gaugino masses.

\subsection{Dirac Gaugino Masses}

Gluino mass contributions via explicit R-violation are not an option since TeV scale explicit R-breaking ruins the PQ-mechanism.
One can look to alleviate the tension between the gaugino masses (in particular, the gluino)  and TeV scale supersymmetry in an R-symmetric manner via further model building. Specifically, the introduction of chiral superfields in the adjoints of each of the gauge groups allows one to form $R$-symmetric Dirac masses for the gauginos \cite{Fayet:1975yi,Hall:1990hq,Fox:2002bu}.
For a chiral superfield $\mb\Phi_g=\phi+\sqrt{2}\vartheta\widetilde{\phi}+\vartheta^2F$ which is an adjoint under SU(3) and carries zero $R$-charge (see Table \ref{Tab2}), one can construct a Dirac mass term for the gluinos. An SU(2) adjoint $\Phi_W$ can be added for the Winos, although as the loop-induced mass of eq.~(\ref{eee}) is relatively large this is not strictly needed. 
A Bino Dirac partner must be a singlet representation, $\Phi_B$, but singlet fields should be avoided as they allow a coupling to the supersymmetry breaking hidden sector $z\Phi_B$,  complicating the mechanism of Section \ref{sec:3}. 

The manner in which Dirac gaugino masses arise depends on how supersymmetry is broken. In Section \ref{sec:3} we discussed breaking supersymmetry, and tuning the cosmological constant to zero with an R-symmetry, this led to two main possibilities
\begin{itemize}
\item[i)] Pure $F$-term  supersymmetry breaking with $F\sim m^2$, see Section \ref{AC3}.
\item[ii)] Mixed $F$ and $D$ terms supersymmetry breaking with $F\sim D\sim m^2$, see Section \ref{3.5}.
\end{itemize}
In both of these cases the gravitino mass is $m_{3/2}\sim m^2/M_{\rm Pl}$ and assuming gravity mediation the scalar masses $m_0$ are related to this scale by $m_{3/2}/m_0\sim \mathcal{O}(1)$.

The most common approach for introducing Dirac gaugino mass terms is via $D$-term supersymmetry breaking operators   of the form \cite{Hall:1990hq,Fox:2002bu}
\beq
\mathcal{L}\supset\int d^2 \vartheta \, \sqrt{2} \, {\mb W'}_\alpha {\mb W}^\alpha_j \frac{{\mb A}_j}{M}~,
\eeq
such that after inserting the $D$-term spurion VEV $\langle {\mb W'}_\alpha \rangle = \vartheta_\alpha D'$, one has $\mathcal{L}\supset m_D\widetilde{g}_a \widetilde{\phi}_a$ with $m_D=D/M$. In the case that  $F$ and $D$-term breaking is comparable, as assumed in  Section \ref{3.5} , the Dirac mass will be $m_D=m^2/M\sim m_{3/2}$. The $D$-term that sources $m_D$ could be associated to U(1)${}_R$ if it is gauged in the UV via the Green-Schwarz mechanism, or $m_D$ could arise from a different U(1) $D$-term. In the case that the $D$-term is associated to the Green-Schwarz mechanism, then the size of $D$ may be bounded, and requiring $m_D\sim m_{3/2}$ implies that U(1)${}_R$ must be weakly gauged in the UV. We discuss this further in Section \ref{axdecay}.

Even in the absence of $D$-term breaking $F$-term Dirac mass operators can be formed  \cite{Amigo:2008rc}
\beq
\mathcal{L} \supset \int d^2 \vartheta \, \frac{1}{M^3} \,  \Phi_a W_{a,\alpha} D^2 D^\alpha (\Xi^\dagger \Xi)~.
\eeq
where $\Xi$ is an F-term supersymmetry breaking spurion.
These operators tend to lead to gaugino masses of order $m_D\sim F^2/M^3\sim m^4/M^3$, which are naively suppressed by a factor of $(m/M)^2$ relative to the scalar sector.
A potentially viable approach towards  $F$-term Dirac gauginos is based on  screening the scalar soft masses \cite{Abel:2011dc} building on models of deconstructed gaugino mediation \cite{Cheng:2001an,Csaki:2001em}.
Additionally, if the Standard Model sfermion masses can be screened then there is no need introduce the ``Wino Dirac Partner''  $\mb{\Phi_W}$, and the radiatively generated electroweakino masses can be comparable to the sfermions masses. This may also allow more flexibility in the choice of the lightest supersymmetric particle, which otherwise is typically the Bino (or possibly the Wino).

\subsection{Proton Decay Operators}

With the imposition of the R-symmetry renormalisable operators that violate $B$ and $L$, leading to fast proton decay, are forbidden by the U(1) symmetry except in a number of special cases. We can calculate the net R-charge of these operators 
\beq
R[\mb{H_uL}] &= 1\neq2~,\\
R[\mb{LQD^c}] &= 3-A\neq2~,\\
R[\mb{LLE^c}] &= 3-A\neq2~,\\
R[\mb{U^cD^cD^c}] &=6-3Q-A(1+\sin^2\beta)\neq2~.
\eeq
The first three operators are automatically absent given our prior restrictions on $A$ (cf.~eq~(\ref{0124})) and the operator is generically forbidden.

Similarly, we can examine the dimension-five operators that violate lepton or baryon number
\beq
R[\mb{LH_uH_uH_d}] &= 1+A\neq2~,\\
R[\mb{QH_dU^cE^c}] &= 3\neq2~,\\
R[\mb{QQQL}] &= 1+3Q-A\cos^2\beta\neq2~,\\
R[\mb{QQQH_d}] &= 3Q+A\sin^2\beta\neq2~,\\
R[\mb{U^cU^cD^cE^c}] &= 7-3Q-A(1+ \sin^2\beta) \neq2~.
\eeq
There is a great amount of freedom for $A$, $Q$, and $\beta$. Thus all of these dimension-five operators are generically forbidden by the U(1)${}_R$  symmetry (especially so if one does not insist on rational R-charges). 

Let us just check one simple example.
Consider the case  $A=1/(3\cos^2\beta)=10/3$, this corresponds to the restricted applied in Table \ref{Tab1} to simplify the charges. There is nothing special about this choice except it is helpful for finding rational values. With $A$ fixed we find that all of the above proton decay operators are forbidden provided $Q\neq5/3,~4/9,-1/3,-7/9$. 

One might be concerned that after spontaneous R-symmetry breaking the model loses the protection of the U(1)${}_R$  leading to dangerous proton decay operators suppressed only by the R-breaking scale. Specifically, operators of the form $\mathcal{O}{\mb X^n \mb Y^m}$, may lead to low dimension proton decay operators after VEV insertions, however this seems to generally not be an issue.
Indeed, for certain charge assignments there is a clear residual discrete R-symmetry which continues to forbid proton decay operator. Most notably for the binary charges assignments (although this is a case we do not consider here) in which the fields ${\mb Q}$, ${\mb U^c}$, ${\mb D^c}$, ${\mb L}$, ${\mb E^c}$ have R-charge 1 (odd $R$-charge) and ${\mb H_u}$, ${\mb H_d}$, ${\mb X}$, ${\mb Y}$ have R-charge zero (even $R$-charge). In this case only the even R-charged scalars that acquire VEVs, as a result, there is a residual $Z_2$. This $Z_2$ symmetry is the regular $R$-parity $R_p=(-1)^{3B-L+2s}$ in which the quark and lepton superfields are odd and the other fields are even. 

For more general $R$ charges it is difficult to identify the condition for residual discrete symmetries. A comprehensive analysis is prohibitive, however dangerous proton decay operators only arise for special values of $A$, $Q$, and $\tan\beta$. Since the only fermions lighter than the proton carry $L$ (in particular $m_{3/2}\gg m_p$) the proton can only decay if both $B$ and $L$ violation is present. The simplest example of a problematic case arises if $Q=\frac{1}{3}(4-A(1-\sin^2\beta))$ such that the $B$-violating operator $\lambda''{\mb U^c}{\mb D^c}{\mb D^c}$ is present (note this does not constrain $A$) and some operator of the form $\mathcal{O}_{nm}=\mb{(LQD^c) X^n Y^m}$ is also present. The latter operator has R-charge $R[\mathcal{O}_{nm}]=\frac{1}{2} A (n-2-3 m)+2 m+3$ and the condition $R[\mathcal{O}_{nm}]=2$ is hard to satisfy, particularly since we require $A\not\in \mathbb{Z}$. Once the $X$ and $Y$ get VEVs this implies $\mb{(LQD^c)}$ has an effective coupling $\lambda'\sim (f_a/M_{\rm Pl})^d\sim10^{-8d}$ where $d=n+m$. For example take $A=5/3$ such that the operator $\mathcal{O}_{22}$ is allowed (but no lower dimension operators), thus $d=4$ and this implies a proton lifetime of order 
\beq
\tau_p
&\sim\left(\frac{(\lambda'\lambda'')^2 m_p^5}{m_{\rm soft^4}}\right)^{-1}\\
&\sim
10^{27}~{\rm s}
\left(\frac{10^{-5}}{\lambda''}\right)^2
\left(\frac{10^{-8}}{f_a/M_{\rm Pl}}\right)^4
\left(\frac{m_{\rm soft}}{3~{\rm TeV}}\right)^4~.
\eeq
The experimental limit is $\tau_p>10^{34}$ s, thus scenarios can be excluded by proton decay for certain model parameters, but this is rare (specifically, here we made a special choice for $Q$).

In contrast to this cautionary example, since the R-charge assignments tend to be distinct between fields, we find that $B$ and $L$ violating operators which are R-invariant tend arise only at relatively high mass dimension. 
It is difficult to make a general statement, other than it would seem that the residual impact of the R-symmetry is to protect against proton decay even after R-breaking (this is reminiscent of the result in \cite{Kribs:2007ac} regarding flavour violation). Moreover, if the $R$ charges of fields are taken to be irrational we anticipate that high dimension operators which lead to observable proton decay will be generically absent.

\subsection{Lightest Supersymmetric Particle}

The typical absence of low dimension R-violating operators also implies that the lightest supersymmetric particle (LSP) will be metastable. From the model building considerations above, the LSP is most likely the Bino or Wino. Such electroweakino LSPs are fermions and thus their decay must involve a Standard Model fermion. Since all of the Standard Model fermions carry $B$ and $L$ this implies LSP decays must involve a $B$ or $L$ violating operator which we have seen are highly suppressed even after spontaneous R-breaking.

We can estimate the lifetime of the electroweakino LSP assuming it decays via a dimension four operator suppressed by powers of $( f_a/M_{\rm Pl})^d\sim10^{-8d}$, similar to the example above. The LSP lifetime is parametrically
\beq
\tau_\chi\sim \left[m_{\rm LSP}\left(\frac{f_a}{M_{\rm Pl}}\right)^{2d}\right]^{-1} \sim 10^{16d-27} ~s~,
\eeq
where $d$ corresponds to the number of VEV insertions (of $X$ or $Y$) needed to generate the dimension four R-partity violating operator. For $d\geq3$ (and $f_a\sim10^{10}$ GeV) the LSP lifetime should be longer than the age of the Universe (i.e.~$\tau_\chi\gtrsim10^{18}$ s); with $d=3$ this corresponds to $\tau_\chi|_{d=3}\sim10^{21}$s. We have argued above that $B$, $L$ violating operator with  low mass dimension are rare, thus we expect that for generic choices for $A$ that the LSP will be metastable. 

It is interesting to note that for $d\lesssim3$ the LSP lifetime is typically $\tau_\chi\ll1$s which allows the LSP to decay before any observable limits apply. Such a scenario would arise without proton decay, if there is $B$ violation is modest, while $L$ violation is heavily suppressed, or vice-versa.
Furthermore, Binos tend to be poor thermal dark matter candidates, but overproduction can be avoided via R-violating decays or in the case of a low inflationary reheat temperature.
An interesting example of an unstable LSP is found for $A=6/5$ in this case the $L$-violating operators $\lambda'\mb{(LQD^c) Y}$ and $\lambda \mb{(LLE^c) Y}$ appear as ($n=1$) higher terms in the superpotential. Inserting the $X$ and $Y$  VEVs  leads to effective couplings $\lambda,\lambda'\sim (f_a/M_{\rm Pl})\sim 10^{-8}$ and thus $\tau_\chi\sim10^{-11}$ s.
Conversely, a cautionary example is given by $A=3/2$ for which $\tau_\chi\sim10^{5}$ s and such late decays may be in tension with cosmological observables.

An alternative candidate for the LSP can actually be found within the axion supermultiplet, which contains the fermion axino and complex scalar saxion fields. The axino and saxion will typically have masses of order the superpartner mass scale, although the details depend on the specifics of supersymmetry breaking. An axino LSP can potentially be a viable dark matter candidate  \cite{Covi:1999ty} (also see review \cite{Choi:2013lwa}).

For gravity mediation all superpartner states are of the same characteristic mass scale. However, the details of the UV completion into string theory can lead to small splittings between the states. As a result, the gravitino may not be the LSP, and the axino can potentially be the LSP, see in particular \cite{Kim:1983ia,Rajagopal:1990yx,Chun:1992zk,Goto:1991gq}
Typically in gauge-mediated supersymmetry breaking (or similar) the gravitino will generically be the LSP. However, there exist supersymmetry mediation mechanisms that lead to an axino LSP, see for instance: \cite{Nakamura:2008ey,Colucci:2018yaq,Bae:2020hys}.

It is also worth highlighting that the saxion can have interesting dynamics that can further complicate the cosmological history of dark matter in supersymmetric axion models. Notably, if the saxion forms a condensate in the early Universe, this potentially can dilute any dark matter abundance \cite{Co:2017orl}.

The relic density commonly depends on the full spectrum of the supersymmetric particles. The superpartner spectrum depends sensitively on the details of supersymmetry breaking, which offers a great degree of model freedom. This allows for rich dark matter phenomenology, and a full analysis of the range of dark matter phenomenology is beyond the scope of the current paper.

\section{String-inspired Axions}
\label{sec:5}

Global symmetries are thought to be explicitly violated by Planck scale physics \cite{Banks:2010zn}, which is potentially ruinous for the PQ mechanism. It is possible that Planck scale violations do not have $\mathcal{O}$(1) coefficients, but rather are exponential suppressed \cite{Kallosh:1995hi}, in  which case  explicit breakings due to gravity would be non-issue. Setting aside this caveat, if one parameterizes the symmetry violating Planck scale physics in terms of $(1/M_{\rm Pl})^{n}$-suppressed operators, one requires all operators of mass-dimension 9 or less to be absent  \cite{Kamionkowski:1992mf,Holman:1992us,Barr:1992qq}. While these bothersome operators can be removed through new discrete or continuous gauge symmetries \citep{Georgi:1981pu,Bhattiprolu:2021rrj,Randall:1992ut,Redi:2016esr,DiLuzio:2017tjx,Fukuda:2017ylt,Choi:2022fha,Duerr:2017amf}, UV completions into string theory and supergravity provide an alternative resolution.

 In many string compactifications the low-energy theory contains anomalous, abelian U(1)${}_A$ symmetries which can be gauged in the UV via the Green-Schwarz (GS) mechanism \cite{Svrcek:2006yi,Green:1984sg}. 
The prospect of a identifying the PQ symmetry with an anomalous global remnant of the GS mechanism has been previously explored in a variety of string models, see e.g.~\cite{Heckman:2008qt,Arvanitaki:2009fg,Demirtas:2021gsq,Buchbinder:2014qca,Choi:2014uaa,Ahn:2016typ,Conlon:2006tq}.
Our motivation for considering the GS mechanism is thus twofold: we should like to identify a UV completion of the anomalous U(1)${}_{PQ}$ compatible with string theory, and by gauging the R-symmetry we ensure that the axion is protected against $M_{\rm Pl}$-induced R-violating operators.

From a ``bottom-up'' phenomenological perspective the GS mechanism  introduces a number of restrictions. In particular, consistent implementation of GS implies an anomaly constraint (specifically the field content must be such that there is no U$(1)_{Y}\times $U$(1)_R^2$  anomaly in addition to the GS cancellation conditions), a restriction on the size of the $D$-term associated to U(1)${}_R$ (and thereby $f_a$), and it imposes conditions on the structure of the UV  completion into string theory. We discuss these in detail shortly.
Since the GS mechanism is linked to string theory one may have concerns related to moduli stabilisation (in particular those moduli involved in the implementation of GS). Many issues relating to moduli stabilisation remain unresolved. There also may be subtleties relating to the fact that we have assumed in Section \ref{sec:3} that neither the K\"ahler or superpotential introduce explicit R-breaking terms, which may lead to deviations from standard approaches to stabilising moduli. We do not attempt to address these issues, and do not attempt to provide a detailed UV-complete model, as they are beyond the scope of this work. Rather, we will assume that such issues can be solved and explore the novelties involved with implementing the GS mechanism to gauge U(1)$_{R}$.

\subsection{Green-Schwarz Mechanism}\label{gs_sec}

There are a variety of axions candidates provided by string theory. In particular, there is the model-independent axion, where the role of the axion is played by the imaginary part of the dilaton superfield, and model-dependent axions given by the imaginary parts of the K\"ahler, volume, or structure moduli. 
In Appendix \ref{app-string}, we outline how  axions can arise from the closed string sector (see e.g.~\cite{Svrcek:2006yi} for more complete discussions). We note that such states tend to suffer from too large a decay constant to be  viable candidates for resolving the strong CP problem\footnote{Many string models have been constructed in which more favourable axion decay constant are obtained, for example for model-dependent axions in the LARGE volume scenario \cite{Balasubramanian:2005zx,Conlon:2006tq}.}. 
Axions with more appropriate decay constants may be obtained by a mixing of the closed string axion with open string axions arising from the gauging of an anomalous U(1)${}_A$ symmetry via the GS mechanism \cite{Green:1984sg}, as we discuss below.

 In string theory gauge anomalies can be cancelled by the GS mechanism by allowing U(1)${}_A$ to act on the modulus superfield $T=\tau + i\theta_{\rm s}$ as a shift symmetry
 \beq \label{define}
{\rm U}(1)_A: \qquad V_A &\rightarrow V_A+\frac{i}{2}(\Lambda- \overline{\Lambda}), \\
T &\rightarrow T+i\delta_{\rm GS}\Lambda, \\
\phi_j &\rightarrow e^{iq_j\Lambda}\phi_j,
\eeq
where $V_A$ is the vector superfield for the U$(1)_A$ gauge multiplet,
$\phi_j$ are generic U(1)${}_A$-charged chiral matter superfields,
and $\Lambda$ is a chiral superfield parameterizing the U$(1)_A$ transformation on the superspace. Notably, the modulus utilysed to implement the GS mechanism $T$ can be any modulus field i.e.~the dilaton, K\"ahler or volume modulus.

The shift symmetry of eq.~(\ref{define}) is associated to an axion $\theta_{\rm s}$ which can be coupled to Standard Model instantons leading to the topological term proportional to $\theta_{\rm s}$. This coupling allows for the stringy axion $\theta_{\rm s}$ to act as the QCD axion, potentially solving the strong CP problem.\footnote{We will assume that the stringy instanton contributions are small enough so as to avoid quality problems, examples of string models achieving this can be found in e.g.~\cite{Conlon:2006tq,Demirtas:2021gsq}.} Furthermore, if some U(1)${}_A$ charged chiral superfields obtain a VEV, there will be an additional axion-like particle as per the usual PQ mechanism. One linear combination of these axions 
will be eaten, giving a mass to the U(1)${}_A$ gauge field,
spontaneously breaking the gauge symmetry and leaving behind a global anomalous U(1)${}_{\rm PQ}$ symmetry. The other linear combination provides a  QCD axion candidate as will be seen explicitly in Section~\ref{axid}.

In what follows we will focus  our discussion on the GS mechanism in the context of the $\mathcal{N}=1$ $D=4$ supergravity Lagrangian without further reference to UV string completions.
One finds (see Appendix \ref{app-string}) that the relevant part of the Lagrangian is given by 

\beq \label{fullL}
{\cal L} =&-\frac{1}{4g^{2}_{i}}G_{i}^{\mu\nu}G_{\mu\nu,i}+ K_{T\overline{T}}\left(\partial_\mu \theta_{\rm s}- \delta_{\rm GS}A_\mu\right)^2
+ D_\mu \phi^*_{j} D^\mu \phi_{j}\\
&+\frac{g^{2}_R}{2}\left(-\delta_{\rm GS} K_{T} -\sum_{j}q_{j}|\phi_{j}|^2 +\xi_{R}\right)^2\\
&+\frac{1}{4} k_{i}\theta_{\rm s} G_{i}^{\mu\nu}\tilde G_{\mu\nu,i}+{\cal L}_{\rm Matter}+\cdots
\eeq
where $k_{i}g^2_{i}=g^2_{\rm GUT}$, the unified GUT coupling, $i$ labels the gauge group and $g_R$ is the coupling of the local R-symmetry.

Since we have identified the PQ symmetry as the R-symmetry, we consider here the prospect of gauging U(1)${}_R$ via GS. 
One can see the GS mechanism can cancel a U(1)${}_R$ gauge anomaly from eq.~(\ref{define})~\&~(\ref{fullL}); the anomalous contribution to the Lagrangian due to U(1)${}_R$ transformation is 
\beq
\Delta {\cal L}_{\rm eff}^{\rm anom} = - \frac{\mathcal{A}_{i}}{16\pi^2}G^{\mu\nu}_{i}\tilde G_{\mu\nu,i}~,
\eeq
where $\mathcal{A}_{i}$ denotes the anomaly coefficient  of gauge group labelled by $i$. Under the same U(1)${}_R$ transformation the axion shift leads to
\beq
\Delta {\cal L}_{\rm eff}^{\rm GS} = \frac{\delta_{\rm GS}}{4}k_{i}G^{\mu\nu}_{i}\tilde G_{\mu\nu,i}.
\eeq
Requiring the total Lagrangian be invariant (and hence the anomaly is cancelled) results in the GS anomaly cancellation condition
\beq
\label{GS}
\delta_{\rm GS}=\frac{\mathcal{A}_i}{4\pi^2k_{i}}=\frac{1}{4\pi^2k_{i}}\sum_j q_j {\rm Tr}[T_a^2(\phi_j)].
\eeq
One might worry that this field implementation of the GS mechanism for the purpose of gauging an R-symmetry is incomplete. Remarkably, in \cite{Antoniadis:2014iea} it is shown that this application of the GS  mechanism within this field theoretic approach to the R-symmetry yields the same anomaly cancellation conditions as a full supergravity analysis. 

Note that the $\phi_j$ transform linearly under U(1)${}_R$, whereas the $\theta_s$ transforms as a shift symmetry (i.e.~non-linearly) under the U(1)${}_R$. In the GS mechanism such a shift symmetry is not associated to spontaneous symmetry breaking of some other symmetry but is connected to dimensional reduction from the higher dimensional spacetime (as discussed in Appendix C). 
The U(1)${}_R$ shift symmetry of the GS superfield and the U(1)${}_R$ transformation of the $\phi_j$ cannot be viewed as two separate symmetries. Mainly, as an R-symmetry is an automorphism of the supersymmetry algebra, and both the GS sector and the $\phi_j$ form representations of the same supersymmetry algebra, there is only one U(1)${}_R$. Further, only the combination of the two transformations, made anomaly-free by the GS mechanism, is a sensible symmetry.

An important difference between the R-symmetry and an arbitary U(1) is that  in supergravity gauged R-symmetries are closely tied to constant FI $D$-terms. We review this connection, and some key results, in Appendix \ref{APBBB}. Given that the superpotential transforms as $W \rightarrow e^{i\xi \Lambda} W$ under U(1)${}_R$ the corresponding constant FI term contribution to the $D$-term potential is given by (see eq.~(\ref{A4})) 
\beq\label{kapparel0}
\xi_{R}=-M^{2}_{\rm Pl}\xi ~.
\eeq
Additionally, there is an inherent contribution to the $D$-term from the GS mechanism due to the stabilisation of the GS modulus field $T$. From the general expression eq.~(\ref{dterm}) for the $D$-term potential and the transformation of $T$ under U(1)${}_R$ in eq.~(\ref{define}), it follow that
\beq\label{DDD3}
D_{R} \supset iK_{j}X^{j}_{R} \supset iK_{T}X^{T}_{R}~,
\eeq
where $X^{j}_{R}$ are the supergravity Killing vector fields. Upon stabilisation of the GS modulus $T$ this term induces a so-called \textit{field dependent} FI term. As will be discussed throughout this section, this field dependent FI term has important phenomenological consequences due to the role of $D$-terms in supersymmetry breaking. The gauging of U(1)${}_R$ via the GS mechanism provides an interesting deviation from usual  phenomenology involving GS due to the additional contribution (\ref{kapparel0}) to the $D$-term.

\subsection{Identifying the Axion}\label{axid}

One can identify the axion degree of freedom from inspection of the Lagrangian of eq.~(\ref{fullL}).
The Lagrangian contribution ${\cal L}_{\rm Matter}$ includes a potential for the R-charged scalars  $\phi_{j} $ that gives rise to nonzero VEVs $v_{j}=\sqrt2\langle \phi_{j} \rangle$ for some of these fields. Following \cite{Choi:2014uaa}, we expand the R-charged scalars around these VEVs to obtain
\beq \label{chiral}
\phi_{j}= \frac{v_{j} +\rho(x)}{\sqrt{2}}\exp\left[\frac{i\theta_{j}}{v_{j}}\right]~.
\eeq
As usual in axion models, performing an anomalous field redefinition after the scalars obtain a VEV will generate the axion term
\beq
{\cal L}_{\rm Axion} \supset \frac{1}{4} k_{i}\left(\theta_{\rm s}
- \delta_{\rm GS}\frac{\sum_{j} q_{j}v_{j}\theta_{j}}{\sum_{j} q^{2}_{j}v^{2}_{j}}\right)G_{i}^{\mu\nu}\tilde G_{\mu\nu,i}~.
\eeq
Following this field redefinition U(1)${}_R$ is realised non-linearly as the axionic shift symmetry
\beq
\theta_{\rm s} \rightarrow \theta_{\rm s} + \delta_{\rm GS} \Lambda
\qquad~{\rm and}~\qquad
\theta_{j} \rightarrow \theta_{j} + q_{j}v_{j} \Lambda~.
\eeq
We take the effective Lagrangian of  eq.~(\ref{fullL}), with $\xi_{R}=-M^{2}_{\rm Pl}\xi$ from eq.~(\ref{kapparel0}), and introduce the degrees of freedom $\eta$ and $a$ via field redefinitions
\beq \label{Lfull}
{\cal L}_{\rm eff}=&
-\frac{1}{4g^{2}_{i}}G_{i}^{\mu\nu}G_{\mu\nu,i}\\
&+\frac{1}{2}
\left((8\pi^2\delta_{\rm GS} f_{\rm s})^2+{\hat v}^2 \right)
\Big( \frac{\partial_\mu \eta}{\sqrt{ (8\pi^2\delta_{\rm GS} f_{\rm s})^2 +{\hat v}^2}}-A_\mu\Big)^2
 \\
& +\,
\frac{1}{2}(\partial_\mu a)^2
+ \frac{1}{32\pi^2} \frac{a}{f_a}\,G\tilde G\\
&+\frac{1}{2} g^2_R\left(\delta_{\rm GS} K_{T} + 
\sum_{j}q_{j}|\phi_{j}|^{2}+M^{2}_{\rm Pl}\xi \right)^2 + {\cal L}_{\rm Matter}~,
\eeq
where ${\hat v}^{2}= \sum_{j} q^{2}_{j}v^{2}_{j}$ and we define the linear combinations 
\beq \label{eta}
\eta &=
\frac{1}{\sqrt{(8\pi^2\delta_{\rm GS} f_{\rm s})^2+{\hat v}^2}}
\left(
(8\pi^2\delta_{\rm GS} f_{\rm s})^2 \frac{\theta_{\rm s} }{\delta_{\rm GS}}
+ \sum_{j} q_{j}v_{j}\theta_{j} \right),\\
a &=
\frac{(8\pi^2\delta_{\rm GS} f_{\rm s}) {\hat v}}{\sqrt{(8\pi^2\delta_{\rm GS} f_{\rm s})^2+{\hat v}^2}}
\left(\frac{\theta_{\rm s} }{\delta_{\rm GS}}- \frac{\sum_{j} q_{j}v_{j}\theta_{j}}{\sum_{j} q^{2}_{j}v^{2}_{j}}\right)~,
\eeq
with $f_{\rm s}=\frac{1}{8\pi^2}\sqrt{2K_{T\overline T}}$  for the closed string axion (see eq.~(\ref{fcl})  in Appendix \ref{app-string}) and 
\beq \label{axiondecay}
f_a &\equiv&
\frac{f_{\rm s} {\hat v}}{k_{3}\sqrt{(8\pi^2\delta_{\rm GS} f_{\rm s})^2+{\hat v}^2}}~.
\eeq
Furthermore, the linear combination $\eta$ forms the longitudinal component of the U(1)${}_R$ gauge boson, giving it a mass
\beq \label{mR}
M_R = g_R \sqrt{(8\pi^2\delta_{\rm GS} f_{\rm s})^2 + {\hat v}^2}~,
\eeq
Conversely, $a$ remains perturbatively massless. According to \textit{Vafa-Witten} theorem \cite{vafawitten}, the axion will dynamically relax to $\langle a \rangle = 0$, thereby solving the strong CP problem.

Thus the axion decay constant $f_a$ is a function of  $\hat v$, $f_s$, and $\delta_{\rm GS}$. One can observe from eq.~(\ref{axiondecay}) that the scalar VEVs $\langle \phi_f\rangle$ partially determine the scale of PQ breaking, however this also depends on the details of the UV model (through $f_s$ and $\delta_{\rm GS}$). 
Specifically, if $f_s \gg {\hat v}$ it follows from eq.~(\ref{axiondecay}) that $f_a \sim {\hat v}/(8\pi^{2}k_3\delta_{\rm GS}) \sim {\hat v}$. To make the connection with our models of Section \ref{sec:2} these ${\hat v}$ should be identified with the visible sector R-breaking VEVs of the $X$ and $Y$ scalars. Moreover, using eq.~(\ref{fcl}), one has $M_{R} \sim g_{R}\delta_{GS}M_{\rm Pl}$. Hence in this regime $f_a$ is set by the PQ-charged scalar VEVs. 
We will discuss the conditions under which $f_a\ll M_{\rm Pl}$ in greater detail in Section \ref{axdecay}.

\subsection{U(1)${}_R$ and Field-Dependent FI Terms}\label{5.2}

Before proceeding, we highlight an important subtlety involved in gauging U(1)${}_R$ that is often overlooked. The superpotential transforms as $W \rightarrow e^{i\xi\Lambda(x)}W$ under a gauged U(1)${}_R$ and thus the Lagrangian of eq.~(\ref{fullL}) includes a constant FI $D$-term contribution $\xi_{R}=-M^{2}_{\rm Pl}\xi$. Conversely, including a constant FI $D$-term for a gauged vector multiplet results in a Lagrangian that is invariant  under the gauge symmetry only if the superpotential transforms non-trivially (see Appendix.~\ref{superconformal}). 

Even the gauging of an initially non-$R$ anomalous U(1)${}_A$ via the GS mechanism results in a FI-like contribution to the $D$-term of the form of eq.~(\ref{DDD3}). Hence is it natural to ask whether, after stabilisation of the GS modulus, the resulting effective superpotential transforms non-trivally under such a U(1)${}_A$. More precisely, from eq.~(\ref{DDD3}) it can be seen that the  U(1)${}_A$ shift of the GS modulus, given in eq.~(\ref{define}), provides a contribution to the $D$-term potential
\beq
D\supset iK_{T}X^{T}_{R}=-K_{T}\delta_{\rm GS} \equiv \xi_{\rm GS}.
\eeq 
If $K_T$ obtains a nonzero VEV this contribution closely resembles  a constant FI $D$-term contribution. The $\xi_{\rm GS}$ term is referred to as (loop-induced) \textit{field dependent} FI term and arises whenever a U(1) symmetry is gauged via the GS mechanism \cite{Dreiner:2003yr,Dreiner:2003hw,Dreiner:2013ala,Antoniadis:2014iea,Svrcek:2006yi}.  Thus it is relevant for phenomenology to establish whether the constant FI term and field-dependent FI term combine into an effective FI term $\xi_{\rm eff} = -M^{-2}_{\rm Pl}(\xi_{\rm GS}+ \xi_R)$ resulting in $W\rightarrow e^{i\xi_{\rm eff}\Lambda} W$. This question is discussed in detail in \cite{Binetruy:2004hh,Elvang:2006jk,Binetruy:1996u}. The conclusion is that if the GS modulus can be stabilized whilst preserving supersymmetry the low energy effective superpotential can indeed transform as $W \rightarrow e^{i\xi_{\rm eff}\Lambda}W$, depending on the form of the K\"ahler and superpotential used to stabilise  the GS moduli (see, in particular, Sections 3.4 \& 6 of \cite{Binetruy:2004hh}). 

As seen in Section \ref{axid}, the imaginary part of the scalar component of $T$ denoted by $\theta_s$ gets a mass (\ref{mR}) via the Higgs mechanism by forming the longitudinal component of the U(1)${}_A$ vector multiplet. Hence in supersymmetric stabilisation of $T$ the real and imaginary scalar parts of $T$ will have to acquire the same mass, and the entire vector multiplet will be integrated out below this scale. Hence below this scale there will be no $D$-term, but there will be a remnant global anomalous U(1)${}_R$ symmetry under which the effective low energy superpotential transforms as $W \rightarrow e^{i\xi_{\rm eff}\Lambda}W$.
In this case the GS gauging of U(1)${}_A$ will result in an effective global U(1)${}_R$ and hence the model building constraints of previous sections will be relevant even when the symmetry gauged by GS is not an R-symmetry. For non supersymmetry-preserving stabilisation of $T$ however the field dependent FI term does not contribute to the transformation of the superpotential, implying 
\beq \label{FIterm}
D_{R}&\supset  \xi_{\rm GS}+ \xi_R~, \\
 W&\rightarrow e^{-iM^{-2}_{\rm Pl}\xi_{R}\Lambda} W.
\eeq

In this work we are interested in a global U$(1)_R$ PQ symmetry that is gauged in the UV via the GS mechanism. The above discussion highlights that, depending on the details of how $T$ is stabilised, this global U$(1)_R$ can be the remnant of either a gauged R-symmetry or an ordinary U$(1)_A$. The key phenomenological difference between these two possibilities is the contributions to the $D$-term, with the gauged R-symmetry having the extra contribution of eq.~(\ref{kapparel0}). We will discuss the implications of this constant FI $D$-term contribution below.

\subsection{Phenomenological Implications of the Green Schwarz Mechanism} 
\label{axdecay}

Application of the GS mechanism to a U(1)${}_A$ symmetry implies the following form for the associated $D$-term potential
\beq \label{eq:nocfi}
V_{D}=\frac{g^2_A}{2}\left(-M^{2}_{\rm Pl}\delta_{\rm GS} K_{0,T} + \sum_{j}q_{j}|\phi_{j}|^2\right)^2~,
\eeq
where $K_0$ is the K\"ahler potential of the matter fields, we will not comment on its exact form (although see Appendix \ref{app-string}) but will discuss its expected magnitude.
As discussed in \cite{Kawamura:1998gy,Choi:2006bh}, to obtain viable low energy phenomenology with $m_{3/2} \ll M_{A}$ with $M_A \sim M_{\rm Pl}$ and $\langle V_{\rm scalar}\rangle \approx 0$ the $D$-term must satisfy the following bound\footnote{This bound follows from combining the stationary condition $\partial_i V =0$ with the transformation of the potential $X^{i}_{A}\partial_i V$  under U(1)${}_A$.}
\beq \label{bound}
D_A=\left(-M^{2}_{\rm Pl}\delta_{\rm GS} K_{0,T} + \sum_{j}q_{j}|\phi_{j}|^2\right)
~\lesssim~ \frac{m^{2}_{3/2}M^{2}_{\rm Pl}}{M^{2}_{R}} ~.
\eeq

The bound of eq.~(\ref{bound}) can restrict the scale of $f_a$. Let us first discuss ``non-R" anomalous U(1)${}_A$ symmetry (i.e.~U(1)${}_A \neq$ U(1)${}_R$) (thus without a constant FI $D$-term) gauged by the GS mechanism. Equation~(\ref{mR}) gives the mass of the U(1)${}_A$ vector boson. Suppose that the GS modulus $T$ is the dilaton or a K\"ahler modulus then $K_{0,T}$ is expected to be $\mathcal{O}(1)$ \cite{Svrcek:2006yi,Choi:2014uaa}.
 Thus to satisfy eq.~(\ref{bound}) one of the PQ charged scalars must obtain a Planck scale VEV $\hat{v}\sim \langle\phi\rangle \sim M_{\rm Pl}$ implying $f_a \sim f_s \sim M_{\rm Pl}$ which is undesirable. 
However, if instead $T$ is stabilised at a tuned point near that of vanishing $\xi_{\rm GS}$ then $|K_{0,T}|\ll 1$, and the bound of eq.~(\ref{bound}) can be satisfied by tuning $|K_{0,T}| < \langle\phi\rangle^2/\delta_{\rm GS}M^2_{\rm Pl}$. Notably, this maintains the freedom to choose the scale of $f_a\ll M_{\rm Pl}$. Hence in this class of models the GS mechanism can be used to obtain an axion with a desirable $f_a$  (see \cite{Conlon:2006tq,Kim:2006aq,Cicoli:2013cha,Buchbinder:2014qca} for examples of this is in Type II and heterotic string theory). 

The case in which U(1)${}_R$ is gauged via the GS mechanism is different since the $D$-term also has a contribution from the constant FI term, defined by $W\rightarrow e^{i\xi\Lambda}W$, thus
\beq \label{fullD}
 V_D =\frac{g^2_A}{2}\left(-M^{2}_{\rm Pl}\delta_{\rm GS} K_{0,T} + \sum_{j}q_{j}|\phi_{j}|^2+M^{2}_{\rm Pl}\xi\right )^2~.
\eeq
The bound in eq.~(\ref{bound}) still holds in the presence of a constant FI $D$-term as can be checked by explicit calculation. In order to satisfy eq.~(\ref{bound}) while maintaining freedom for the scale $f_a$ in this case requires 
\beq \label{Rbound}
M^2_{\rm Pl}(\delta_{\rm GS}K_{0,T} -\xi) - \langle\phi\rangle^2 \lesssim\frac{m^{2}_{3/2}M^{2}_{\rm Pl}}{M^{2}_{R}} ~.
\eeq

In a similar argument to the non-R case above, it is desirable to satisfy the bound (\ref{Rbound}) without forcing $\hat{v} \sim M_{\rm Pl}$ such that one can maintain $f_a\ll M_{\rm Pl}$. Anticipating that $\hat{v}^2 \sim f_{a}^2 \sim m_{3/2}M_{\rm Pl}$ the bound of eq.~(\ref{Rbound}) requires a tuning between the FI terms and the VEV $\hat{v}\sim \langle\phi\rangle$. At this stage, $\xi$ is a nonzero but otherwise unconstrained\footnote{See the end of Section~\ref{sec:cancel} for a discussion on the potential restrictions on $\xi$.}
constant put into the theory by requiring a gauged U(1)${}_R$ symmetry under which $W \rightarrow e^{i\xi\Lambda(x)}W$. Thus $\xi$ could be tuned  to satisfy eq.~(\ref{Rbound}). This would allow $f_a$ to be an undetermined scale even for generic values of $K_{0,T} \sim \mathcal{O}(1)$, but has the disadvantage of requiring \textit{ad hoc} tuning of $\xi$. 

As $\delta_{\rm GS}$ is fixed by the anomaly cancellation condition eq.~(\ref{GS}) and the gauge anomalies depend on the R-charges of the fermions, which in turn depend on $\xi$ (see eq.~(\ref{fcharge})), this implies $\delta_{\rm GS}=\delta_{\rm GS}(\xi,q_{i},k_{A})$. Therefore given the value of $K_{0,T}$ at the point of moduli stabilisation, we can choose $\xi$ so as to satisfy eq.~(\ref{Rbound}).
Alternatively, since typically $\xi\sim \mathcal{A}_i$ by eq.~(\ref{GS}) we have that $\delta_{GS} \sim \xi/\pi^2$, and it is possible that $T$ is stabilized at a point for which $K_{0,T}\sim\mathcal{O}(1)$ is sufficiently closely to $\xi/\delta_{\rm GS}$ such that eq.~(\ref{Rbound}) is satisfied. A third option is that the $D$-term bound is satisfied by a hidden sector R-charged field obtaining a Planck scale VEV. In particular, if the GS modulus is stabilised while preserving supersymmetry one might hope to employ the model of Section \ref{AC3} to break supersymmetry and tune the overall cosmological constant to zero. As per the discussion regarding hidden sector contributions to $f_a$ in Section \ref{se3.4}, this field would not give the standard contribution $\sum_{j} q^{2}_{j}v^{2}_{j}$ to $\hat{v}$, instead its contribution will be highly suppressed. In the case that U$(1)_A \neq {\rm U}(1)_R$ this does not hold, as the decoupling of the hidden sector results in two separate U$(1)$ symmetries. However for U$(1)_R$ this is not the case since $\theta$ transforms in both sectors. We leave the details of R-symmetric supersymmetry preserving stabilisation of $T$ with the addition of the supersymmetry breaking R-symmetric hidden sector with non-trivial K\"ahler potential to future work.

If eq.~(\ref{Rbound}) is satisfied the VEV ${\hat v}$ will be determined by the potential $V(\phi_i)$ used to implement the KSVZ/DFSZ model which is constructed such that  $ {\hat v}\ll f_{\rm s}$ and hence results in an acceptable decay constant
\beq
f_a \sim \frac{\langle\phi\rangle}{8\pi^{2}k_3\delta_{\rm GS}}~.
\eeq
From eq.~(\ref{eta}) it is clear that when ${\hat v}\ll f_{\rm s}$, $\eta$ consist mostly of $\theta_{\rm s}$ and $a$ mostly given by the phases of the chiral superfields eq.~(\ref{chiral}). In this regime the GS modulus decouples at the scale eq.~(\ref{mR}) leaving behind a remaining global anomalous U(1)${}_R$ PQ symmetry.

In addition to determining the nature of the above fine tuning options, the stabilisation mechanism for the GS modulus is also important for the structure of the soft scalar Lagrangian and the potential explicit breaking the remnant global U(1)${}_R$. In general, a nonzero $D$-term will contribute to the soft scalar masses, and the precise form of such corrections will depend on the stabilisation of the GS modulus. We will leave a detailed study of structure of the soft scalar masses for the case of U(1)${}_R$ gauged by the GS mechanism with explicit stabilisation of the GS modulus to future work. Here we will focus instead on the restrictions that the bound on the $D$-term given in eq.~(\ref{Rbound}) places on the model.

On further examination of  eq.~(\ref{bound}) \& (\ref{Rbound})  (which we rewrite using eq.~(\ref{mR})) 
\beq
D<\frac{m_{3/2}^2M_{\rm Pl}^2}{M_A^2}\sim~ \frac{m^{2}_{3/2}}{g_{A}^2\delta^2_{GS}}~.
\eeq
We note that this bound restricts the size of $D$-term connected to the GS mechanism. If these $D$-terms source the Dirac gaugino masses, then this would restrict the size of these mass terms. Recall from eq.~(\ref{HallG}) that $m_{3/2}\sim m^2/M_{\rm Pl}$ (where $m$ was the hidden sector mass scale) and it follows that Dirac mass for the gauginos generated by D-terms associated to U(1) groups gauged by GS are restricted to be 
\beq
\label{bbb2}
m_D\sim\frac{D}{M_{\rm Pl}}\lesssim 
3~{\rm TeV}\left(\frac{m_{3/2}}{5~\rm TeV}\right)^2
\left(\frac{10^{-8}}{g_A}\right)^2~.
\eeq
Observe that the U(1) needs to be quite weakly gauged for the constraint to be satisfied with phenomenologically useful D-terms. 

This implies that if the D-term associated to U(1)${}_R$ sources the Dirac masses it is required to have $g_R\lesssim10^{-8}$. 
Since our Standard Model states carry R-charge, $g_R$ can be very small without conflicting with the weak gravity conjecture \cite{Arkani-Hamed:2006emk}. This conjecture requires that for any U(1) there exists a state of mass $m$ and charge $q$ such that $g q M_{\rm Pl}>m$. For our case the neutrinos generically carry  $\mathcal{O}(1)$ R-charge, and since the maximum allowed mass of the lightest neutrino is $m_\nu\sim 0.1$ eV this conservatively restricts $g_R\gtrsim 10^{-28}$. A weakly gauged U(1)${}_R$ implies $M_{R}\sim10^{10}~(g_R/10^{-8}) M_{\rm Pl}\ll M_{\rm Pl}$. Effective field theories of global supersymmetry will breakdown above $M_{R}$.
Note, however, that if one insists that gauge coupling unification should also involve the U(1)${}_R$ coupling $g_R\sim g_{\rm GUT}$ (as may be the case if this relation follows from a simple gauge kinetic function), then clearly $g_R\sim10^{-8}$ is untenable.

Alternatively, a different U(1) factor, unrelated U(1)${}_R$ and the GS mechanism, could gives rise to the D-term $\hat D$ responsible for the Dirac gaugino mass such that $m_D\sim (\hat D+D_R)/M_{\rm Pl}$.  In this case one can have $D_R\sim m_{3/2}^2/M_{\rm Pl}\ll m_D$ and one can take $g_R\sim1$ in the bound of eq.~(\ref{bound}) whilst still realising $m_D\sim$ TeV. The tuning of cosmological constant then requires that the F-term scalar potential $V_F$ cancels against the combined $V_D=(\hat D^2+D_R^2)/2$.
We highlight that there exist models with vanishing cosmological constant in models with U(1)${}_R$ gauged via GS, see e.g.~\cite{Antoniadis:2014iea,Antoniadis:2014hfa,Antoniadis:2015mna}.


\subsection{Constraints from the Green-Schwarz Mechanism}\label{sec:cancel}

The Green-Schwarz mechanism poses anomaly constraints on the low energy field theory 
\cite{Dreiner:2003yr,Dreiner:2003hw,Dreiner:2013ala}.
In the case that the Standard Model group $SU(3)_{c}\times SU(2)_{W}\times U(1)_{Y} $ is the remaining IR gauge symmetry the GS mechanism gives the following constraints 
\beq \label{GScon}
\frac{\mathcal{A}_{1}}{k_1} =\frac{\mathcal{A}_{2}}{k_2} = \frac{\mathcal{A}_{3}}{k_3} =\frac{\mathcal{A}_{RRR}}{3k_R} =\frac{\mathcal{A}_{GGR}}{24}= 4\pi ^2\delta _{\rm GS}~,
\eeq
where $\mathcal{A}_N$ denotes the ${\rm SU}(N) \times {\rm SU}(N) \times U(1)_A$ anomaly coefficient, $k_N$ is the Kac-Moody level of the $N$-th gauge group, $\mathcal{A}_{GGR} = {\rm Tr}(Q_R)$ is the gravitational anomaly and $\mathcal{A}_{RRR}= {\rm Tr}(Q_R^3)$. 
Independently the $\mathcal{A}_{RRY}$ anomaly has to cancel in order for U(1)${}_R$ to be gauged. For non-Abelian groups $k_i \in \mathbb{N}$  and  $k_1$ is a model-dependent normalisation factor. Most string models are constructed with $k_2 = k_3=1 $, however to remain agnostic about the details of the UV completion we will not impose this restriction, but simply require $k_2/k_3 \in \mathbb{Q}$.

Since gauge coupling unification is desirable for many string theoretic UV completions, this places a further restriction. The VEV of the dilaton field $T$ sets the GUT coupling constant, for heteroic string theory $\langle T\rangle  \sim 1/g^{2}_{\rm GUT}$ and the gauge kinetic function required for the GS mechanism relates the Kac Moody levels and gauge couplings to the GUT coupling
\beq \label{couplings}
g_{\rm GUT}^2=k_{3}g_{3}^2=k_{2}g_{2}^2=k_{1}g_{1}^2=k_{R}g_{R}^2~.
\eeq
This unification condition follows from the assumption that the gauge kinetic function is universal: $\mathcal{F}_{ab}=\delta_{ab}k_a T$. More elaborate UV frameworks may lead to deviations in this condition, allowing for greater freedom; F-theory GUTs provide an example~\cite{Blumenhagen:2008aw,Heckman:2008qt}.

Given the R-charge assignments of the field content the anomaly cancellation condition eq.~(\ref{GScon}) constrains the ratio $k_1:k_2:k_3$. Furthermore, using this relation and  eq.~(\ref{couplings}) one can determine the value of the weak mixing angle \cite{Ibanez:1992fy} at the unification scale, without assuming any particular GUT group
\beq
\sin^2(\theta_W) = \frac{g_{Y}^2}{g_{W}^2+g_{Y}^2}  = \frac{k_2}{k_1 + k_2}~.
\eeq
For $k_1=(5/3)k_2$ this yields the canonical GUT scale  $\sin^2(\theta_W)=\frac{3}{8}$. This condition can provide a guide to constraining the matter content involved in the cancelling the anomalies.

In order to solve the strong CP problem, the U(1)${}_R$ must have a QCD anomaly $\mathcal{A}_{3} \neq 0$  then it follows from eq.~(\ref{GScon}) that one must also require $\mathcal{A}_{2},\mathcal{A}_{1} \neq 0$. 
Specifically, the anomaly coefficient relating to R-SU($N$)${}^2$ are given by 
\beq \label{cancelcon}
\mathcal{A}_{3}^{R} =  & 3 (1 +R[\mb \Phi _g]) \\
 +& \frac{1}{2}\sum_{i=1}^{3} \Big[ (R[\mb U_{i}^C] -1)+ 2(R[\mb Q_{i}]-1)
+  (R[\mb D_{i}^C]  -1 )\Big]~,\\
\mathcal{A}_{2}^{R} =& 2 (1 + R[\mb \Phi_W]  ) + \frac{1}{2}(  R[\mb H_u]  + R[\mb H_d] -2 )\\
&+ \frac{1}{2}\sum_{i=1}^{3}\Big[3 (R[\mb Q_{i}]-1)  + (R[\mb L_{i}]-1)\Big]~, \eeq
and
\beq\label{cancelcon2}
\mathcal{A}_{1}^{R} =& (R[\mb H_{u}]+R[\mb H_{d}]-2) \\
&+ \sum_{i=1}^{3} \Big[ ~\frac{(R[\mb Q_{i}]-1)}{6} +\frac{4 (R[\mb U_{i}^C]-1)}{3} \\
&+ \frac{(R[\mb D_{i}^C]-1) }{3} +\frac{(R[\mb L_{i}]-1)}{2} + (R[\mb E_{i}^C]-1)\Big]~,
\eeq
where the sums are taken over the three generations of quarks and lepton superfields. For the field content of Tables \ref{Tab1} \& \ref{Tab2}, the anomaly coefficient $\mathcal{A}_{3}^{R}$ (relabelled $N$ to match axion conventions) reduces to eq.~(\ref{agg}). 
The gauginos contribute to the anomalies. In Section \ref{sec:4} we argued that Dirac partners were necessary for the gluino and desirable for the wino. Notably, in the case that Dirac partners are included, their fermion components 
 have opposite R-charge to the corresponding gaugino, leading to  cancellations:~$3 (1 + R[\mb \Phi _g]) = 2 (1 + R[\mb\Phi_W ] ) = 0 $.

We will not concern ourselves with the GS conditions imposed by $\mathcal{A}_{GGR}$ and $\mathcal{A}_{RRR}$ (to which the gravitino also contributes) as these can be satisfied by adding hidden sector R-charged SM singlets. The U$(1)_{Y}\times U(1)_A ^2$  anomaly is not cancelled by the GS mechanism and thus one must take care to ensure that $\mathcal{A}_{YRR}=0$, this implies the following condition
\beq\label{anom}
\mathcal{A}_{YRR} &=  \sum_{i}^{3} \Big[(R[\mb Q_{i}]-1)^2  -2(R[\mb U_{i}^C]-1)^2  + (R[\mb D_{i}^C]-1)^2\\
&~~-(R[\mb L_{i}]-1)^2 + (R[\mb E_{i}^C]-1)^2\Big]\\[5pt]
&~~+ (R[\mb H_{u}]-1)^2 -(R[\mb H_{d}]-1)^2 =0~.
\eeq
One can introduce more freedom to the anomaly constraint by adding multiples of any flavour-symmetry to the R-symmetry, say  U$(1)_B$ and U$(1)_L $. If we then take the PQ symmetry to be  U$(1)_{\rm PQ}= R \oplus \gamma B \oplus \zeta L $, resulting the the below additions to the anomaly 
\beq
\mathcal{A}_{i}^{\rm PQ}=\mathcal{A}_{i}^{R} +\mathcal{A}_{i}^{B}+\mathcal{A}_{i}^{L}~,
\eeq with $\mathcal{A}_{i}^R$ given above while $\mathcal{A}_2^{B}= \frac{N_g }{2} \gamma$ and $\mathcal{A}_2^{L}= \frac{N_g }{2}  \zeta $.

The anomaly conditions in eqns.~(\ref{cancelcon}) \& (\ref{cancelcon2}) must hold above the scale $M_R$ (see eq.~(\ref{mR})), before the axionic component of the GS modulus is integrated-out. Therefore Standard Model vector-like fermions that contribute to these anomalies but decouple at a high scale could participate in the cancellation of these anomalies.
We highlight that
\cite{Chamseddine:1995gb} examined the prospect of gauging U(1)${}_R$ for the MSSM supplemented by additional R-charged Standard Model singlet field content. They concluded that if one imposes the constraint $k_1=(5/3)k_2$ and either $k_3=k_2$, or $k_3=2k_2$, or $k_2=2k_3$ (seen as desirable conditions for string constructions), then there are no rational solutions with two or fewer singlet fields (with some additional reasonable assumptions). This reflects the fact that is it non trivial to construct MSSM-like models satisfying eq.~(\ref{GScon}). However, rational solutions likely exist with sufficient field content and, moreover, as noted previously irrational R-charge assignments are acceptable \cite{Castano:1995ci}, although they would correspond to a non-compact gauge group. The FI term $\xi$ determines the R-charge of $\vartheta$, the superpotential,  gauginos, and the gravitino, as well as the difference in R-charge between fermion and boson components of chiral superfields (see Appendix \ref{superconformal}). As the fermionic charges depend on the FI term, requiring a compact gauge group would require the FI term to be quantized $\xi=2n$, for $n \in \mathbb{Z}$. The FI term must also be quantized ($\xi \sim 2\frac{N}{p} $ where $p,N \in \mathbb{Z}$) if the K\"ahler geometry of the field target space has non trivial topology (i.e. the K\"ahler form is not exact) \cite{Seiberg:2010qd,Distler:2010zg,Hellerman:2010fv}.

In summary the conditions for successfully promoting U(1)${}_R$ from a global symmetry into a gauged R-symmetry via the GS mechanism is the cancellation of anomalies, in particular eq.~(\ref{anom}), the UV model building constraint $k_2/k_3 \in \mathbb{Q}$ and $k_1>0$ , and the bound on the D-term eq.~(\ref{Rbound}). The phenomenological benefit of gauging the R-symmetry is that it forces gravity to respect the symmetry and thus removes any Planck violating operators which could spoil the PQ mechanism. Indeed, the global symmetry becomes a geometric symmetry of superspace. We have addressed the other possible quality problem coming from explicit breakings relating to the cosmological constant in Section \ref{sec:3}. Since in this case the only explicit breaking of the R-symmetry is due to the QCD anomaly one has that the QCD angle exactly vanishes: $\bar{\theta}=0$. Furthermore, note that since U(1)${}_R$ is gauged in the UV theory, $M_{\rm Pl}$-suppressed R-violating operators are forbidden and thus gravitational mediation of supersymmetry breaking to the visible sector will preserve the R-symmetry.

Finally, we highlight that while it is elegant to think that one might UV complete the global U(1)${}_R$ to a local symmetry within supergravity, this is not strictly necessary and one could also consider the possibility that the global R-symmetry is accidental.  In this case, the R-axion associated to  spontaneously broken global U(1)${}_R$ symmetry can still be the QCD axion, however, in place of the Green Schwarz mechanism one should introduce other symmetries that protect the R-symmetry from R-violating $M_{\rm Pl}$ suppressed operators up to at least mass dimension 9 (depending on $f_a$) \cite{Kamionkowski:1992mf,Holman:1992us,Barr:1992qq}.  There are two possibilities, either the global U(1)${}_R$ is an accidental symmetry of a large R-symmetry (perhaps a GUT) or there is an additional non-R U(1) or discrete $Z_N$ symmetries which forbid the problematic operators \cite{Georgi:1981pu,Bhattiprolu:2021rrj,Randall:1992ut,Redi:2016esr,DiLuzio:2017tjx,Fukuda:2017ylt,Choi:2022fha,Duerr:2017amf}. An explicit example of an accidental low energy R-symmetry is given in \cite{Kribs:2010md}.

\section{Concluding Remarks}
\label{sec:con}
In this paper we have considered the viability of identifying the PQ symmetry used to solve the strong CP problem with the R-symmetry unique to supersymmetric theories. The PQ solution of the strong CP problem requires the existence of a  global, anomalous U(1)${}_{\rm PQ}$ symmetry. The introduction of this U(1)${}_{\rm PQ}$ can however be rather \textit{ad hoc}. Mainly, in order to solve the strong CP problem, the PQ symmetry needs to be  anomalous under QCD and be explicitly broken only by QCD instanton effects. Quantum gravity however is thought to violate all global symmetries, leading to quality problems. As only gauge symmetries are thought to be consistent with UV completions into quantum gravity, the global anomalous PQ symmetry seems ill motivated. While the QCD anomaly provides an obstruction to the naive gauging of the PQ symmetry, the Green-Schwarz mechanism provides a potential manner of UV completing into an anomaly-free U(1)${}_R$ gauge symmetry, thus avoiding such quality problems. 

With the leading candidate for a theory of quantum gravity being string theory, and the low energy EFT of string theories being theories of supergravity, it is natural to consider the PQ symmetry in supergravity theories. Notably, supersymmetric theories provide a natural candidate for  U(1)${}_{\rm PQ}$: the R-symmetry. R symmetries are outer automorphisms of the super Poincar\'e  algebra that leave the Poincar\'e subalgebra invariant. The supersymmetry generators transform nontrivially under an R-symmetry, leading to rich model building structures that differ from that of non R-symmetries. In order to protect against quality problems, we have explored the possibility of gauging the R-symmetry using the GS mechanism from a bottom-up perspective. However, R-symmetry can only be gauged in theories of supergravity. Additionally, in supergravity any gauge symmetry acts partly as an R-symmetry and the existence of gauged R-symmetries is closely tied to the presence of constant FI $D$-terms. Gauged R-symmetries are therefore well motivated, and we found that the FI $D$-terms associated with them have interesting phenomenological consequences. 

A key ingredient in implementing the PQ mechanism using U(1)${}_R$ was to allow the fields of the MSSM to carry non-binary R-charges. A number of issues arise that derail approaches with minimal R-charge assignments in which quark and lepton superfields have R-charge 1 and Higgs superfields have R-charge 0. One prominent issue is that our approach to breaking supergravity and tuning the cosmological constant with R-symmetric K\"ahler and superpotentials required the hidden sector field $z$ to carry R-charge 2 as discussed in Section \ref{sec:3}. Notably, if the Higgs superfields have R-charge zero then the operator $zH_uH_d$ connects the visible and hidden sectors and ruins the mechanism without further model building. Another critical issue is that with binary charges and a Dirac partner for the gluino there is no R-QCD anomaly. 
We find that these issues are most readily addressed in models of split/high scale supersymmetry and thus we will present this distinct scenario in a companion paper \cite{future}.  
Generalising away from binary charges we have provided a proof of principle that one can construct viable low energy models in which the PQ symmetry is identified with U(1)${}_R$. In particular we identified the field content required in such models in Tables \ref{Tab1} \& \ref{Tab2} of Section \ref{sec:2} and discussed further phenomenological issues in Section \ref{sec:4}. In Section \ref{sec:3} \& \ref{sec:5} we outlined how to maintain axion quality and ensure $\bar{\theta}=0$.

In our framework there are generically two dark matter candidates, namely the lightest supersymmetric particle (LSP) and the R-axion. The LSP can be made metastable via a generic choice of the R-charges.  For special R-charge assignments the LSP can be unstable after R-breaking (generically, fast proton decay is not an issue, but in rare cases it can occur). 

We also highlight that the LSP may be different if the scalar soft masses are screened relative to the gaugino masses. Such an example was proposed by Abel \& Goodsell \cite{Abel:2011dc}. It would be interesting to investigate whether one could implement the R-symmetric hidden sector of Section \ref{sec:3} in the context of a Brane World model \cite{Anisimov:2001zz} in which the visible and hidden sectors reside on different branes within a higher dimensional bulk space. In such models bulk gravitons communicate the supersymmetry breaking  to the visible brane. Notably, if the gaugino fields propagate further into the bulk then one can implement gaugino mediation \cite{Anisimov:2002az}, potentially leading to the suppression of scalar soft masses in the manner of~\cite{Abel:2011dc}.

While a discussion of the cosmology would be better explored in a dedicated paper, we make a few closing remarks.  The traditional `axion window' for $f_a$ is roughly $10^{9}~{\rm GeV}\lesssim f_a\lesssim10^{12}$ GeV~\citep{Patrignani:2016xqp}.
 We highlight that we can arrange for the visible sector R-breaking scale $f_a$ to be appropriate to give the observed dark matter abundance. Indeed, in Section \ref{sec:2} we constructed $f_a$ such that the VEVs of the $X$ and $Y$ fields  give a cosmologically appropriate $f_a\sim10^{10}$ GeV  and simultaneously generate a TeV scale $\mu$-term. The axion is a good candidate for dark matter. The LSP is likely an electroweakino and can play the role of dark matter, although this may require a non-thermal origin if the LSP is mostly Bino. One might also envisage a mixed dark matter scenario. The range of cosmological scenarios in this class of models is rich and we leave a full exploration for future work. However, we are confident that there is enough model freedom to accommodate constraints from dark matter searches and cosmological bounds.

If the R-symmetry is the PQ symmetry then current experimental efforts to search for the QCD axion could discover the R-axion. The R-axion interacts with gluons with a coupling $g_{agg}\propto N$ (as calculated in eq.~(\ref{agg})), it will also generically have a electromagnetic anomaly $E={\rm Tr}Q_RQ_{\rm em}^2$ and couple to photons with strength $g_{a\gamma\gamma}\propto E/N\neq0$ \cite{Srednicki:1985xd}. One can also calculate the axion coupling to quarks and leptons in the standard model-independent manner outlined in \cite{Srednicki:1985xd}.
Moreover, due to the Kim-Nilles mechanism the value of $f_a$ is appropriate for the R-axion to be dark matter (if one discards the superpotential $W_{\mu}$ of eq.~(\ref{Hitoshi}) a wider range of $f_a$ is possible). 

It is also interesting to consider the inverse probelm: if an axion were to be discovered how would one know whether it was associated to the R-symmetry? The positive identification of superpartners (at any scale) would of course lend more weight to this possibility. Moreover, since $f_a$ is the scale of R-symmetry breaking this will impact the superpartner phenomenology and the identification of Dirac gaugino partners would also provide evidence for an R-symmetry. Optimistically, from the axion coupling one could potentially identify the PQ charges of each low energy fields and from these determine if these were appropriate to realise an R-symmetry.


\section*{Acknowledgements} 
JU is supported by NSF grant PHY-2209998 and is grateful for the hospitality of Queen's College, Oxford and the Rudolf Peierls Centre for Theoretical Physics at the University of Oxford. TY is supported by the Peter Davies Scholarship and wishes to thank them for their continued support. We are grateful to Prateek Agrawal, Alan Barr, Herbi Dreiner, and John March-Russell  for insightful interactions.

\appendix
\newpage

\section{Variant Kim-Nilles Realisations}
\label{ApB}
As highlighted in Bhattiprolu \& Martin \cite{Bhattiprolu:2021rrj} there multiple possible realisations of the Kim-Nilles mechanism, in particular:
\beq
W_1&~=\frac{1}{\Lambda}\left(\mb{XYH_uH_d}+\mb{X^3Y}\right)~,\\[6pt]
W_2&~=\frac{1}{\Lambda}\left(\mb{X^2H_uH_d}+\mb{X^3Y}\right)~,\\[6pt]
W_3&~=\frac{1}{\Lambda}\left(\mb{Y^2H_uH_d}+\mb{X^3Y}\right)~,\\[6pt]
W_4&~=\frac{1}{\Lambda}\left(\mb{X^2H_uH_d}+\mb{X^2Y^2}\right)~.
\eeq
For completeness we comment on the R-charge assignments needed to allow these two terms in the superpotential. $W_1$ corresponds to the model of  \cite{Murayama}, which we used as our standard case in the main body of the text. 
Writing $R[H_uH_d]=A$ 
 the R-charges must be
\beq
&X_1= \frac{A}{2}~,   \qquad\qquad  ~~~~~Y_1= \frac{4-3A}{2}~, \\[6pt]
&X_2= \frac{2-A}{2} ~,  \qquad\qquad  Y_2=  \frac{3A-2}{2}~,\\[6pt]
&X_3= \frac{2+A}{6}~,  \qquad\qquad  Y_3= \frac{2-A}{2}~,\\[6pt]
&X_4= \frac{2-A}{2} ~,  \qquad\qquad  Y_4= \frac{A}{2} ~.
\eeq
For the case $W_1$, this is precisely eq.~(\ref{eq7}).
Variants with additional chiral superfields might also be constructed to allow more freedom.

\section{Brief Overview of Supergravity}\label{APBBB}

In this appendix we provide a very brief overview of some elements of $\mathcal{N}=1$, $D=4$ supergravity, for an authoritative source, readers are directed to e.g.~\cite{Nilles:1983ge}. This is helpful for reference in our calculations in Sections \ref{sec:3}, \ref{sec:4} \& \ref{sec:5}.

\subsection{Scalar potential}

The structure of $\mathcal{N}=1,~D=4$ supergravity with chiral and vector supermulitplets is fixed by the real and gauge invariant K\"ahler function $G$, the killing vectors $X_a$, and the gauge kinetic function $\mathcal{F}_{ab}$. In terms of the K\"ahler potential $K$ and superpotential $W$ the K\"ahler function is given by 
\beq
G =\frac{K}{M^2_{\rm Pl}} + \log \frac{|W|^2}{M^6_{\rm Pl}}~.
\eeq

The Killing vector fields $X_a=X^{i}_{a}\frac{\partial}{\partial \phi^i}$ generate the symmetries of the K\"ahler manifold for the scalar fields $\phi^i$ that are gauged by a vector field. The gauge transformation for a given scalar component $\phi^i$ of chiral superfield $\Phi^i$ is given by $\delta \phi^i = X^{i}_a\epsilon^{a}$ where $\epsilon^{a}$ are real gauge parameters. 
For superfields transforming with charge $q_j$ under a gauged U(1)  linear transformation are given by $X^{j}=iq^{j}\phi^{j}$, while for a non-linear shift symmetry $X^{j}=iq^{j}$.

Similar to as discussed in Section \ref{sec:3}, but now in terms of $G$, the scalar potential from supergravity is given by

\beq\label{scalarpot}
V = V_F + V_D = M^4_{\rm Pl}e^{G}(M^2_{\rm Pl}G_{i}G^{i} -3) + \frac{1}{2}D_{a}D^{a}~,
\eeq
where $G_{i}= \frac{\partial G}{\partial {\phi^{i}}}$, gauge indices `$a$' are raised by the inverse of the real part of the gauge kinetic function $[{\rm Re}[\mathcal{F}]^{-1}]^{ab}$ and $G$ indices are raised using the inverse of the K\"ahler metric $K_{i\overline j}= \frac{\partial K}{\partial \phi^{i}\partial{{\overline{\phi}^{j}}}}$. The killing potentials $D_{a}$ are given by 
\beq \label{dterm}
D_{a} = iM^2_{\rm Pl}G_{j}X^{j}_{a} = iK_{j}X^{j}_{a} + iM^2_{\rm Pl}\frac{W_{j}}{W}X^{j}_{a}~.
\eeq

\subsection{Soft Supersymmetry Breaking Operators} \label{softbreak}

We next recall the explicit form of the $\mathcal{N}=1$, $D=4$ supergravity scalar potential (\ref{scalarpot}) as derived in \cite{Brignole:1997wnc}. Upon insertions of the appropriate VEVs these expressions give the soft supersymmetry breaking terms. In particular, these formulae are employed in deriving some of the results of Sections~\ref{sec:3} \& \ref{sec:4}  .

Given hidden sectors superfields $z_h$ and observable sector superfields $\Phi_i$ we use the following parameterisation of the super and K\"ahler potential

\beq \label{W}
W &= \widehat{W}(z_h) + \frac{1}{2}\mu_{ij}(z_h)\Phi_{i}\Phi_{j} +\frac{1}{6}Y_{ijk}(z_h)\Phi_{i}\Phi_j\Phi_{k} + \cdots\\
K &= \widehat{K}(z_h,\overline{z_h}) + \tilde{K}_{\overline{i}j}(z_h,\overline{z_h})\overline{\Phi^i}\Phi^j \\
&+ \Big [\frac{1}{2}Z_{ij}(z_h,\overline{z}_h)\Phi^i\Phi^j + \rm{ h.c}  \Big] + \cdots
\eeq 
where $\widehat{W}$ and $\widehat{K}$ are the hidden sector super and K\"ahler potentials and involve only the hidden sector fields.
Inserting eqs.~(\ref{W}) into the general expression (\ref{softbreak}) for the scalar potential leads to terms of the from
\beq
V \supset \frac{1}{6}A_{ijk}\Phi_{i}\Phi_{j}\Phi_{k} + \frac{1}{2}B_{ij}\Phi_{i}\Phi_{j} + {\rm h.c}.
\eeq
The $A$ terms are given by
\beq \label{Aterm}
A_{ijk} =&~ \frac{\widehat{W}^{*}}{|\widehat{W}|}
e^{\widehat{K}/2M^2_{\rm Pl}}F^{m} 
\Big[\frac{\widehat{K}_m Y_{ijk}}{M^2_{\rm Pl}} \\
&\hspace*{1cm}
 +\partial_m Y_{ijk}-\left(\tilde{K}^{n\overline{p}}\partial_m \tilde{K}_{\overline{p}i}Y_{njk} 
\cdots \right)\Big]~,
\eeq
where the ellipsis indicates terms with $i\leftrightarrow j$ and $j\leftrightarrow k$.
The $B$ terms are given by
\beq\label{Bterm}
B_{ij} &= \frac{\widehat{W}^*}{|\widehat{W}|}
e^{{\hat K}/2M^2_{\rm Pl}}\Big\{ F^m \Big[ 
\frac{{\hat K}_m \mu_{ij}}{M^2_{\rm Pl}}
+ \partial_m \mu_{ij}
\\ &
-\ \left(
\tilde K^{\delta{\overline{\rho}}}
\partial_m {\tilde K_{\overline{\rho}i}} \mu_{\delta j}
+
\tilde K^{\delta{\overline{\rho}}}
\partial_m {\tilde K_{\overline{\rho}j}} \mu_{\delta i}
\right)\Big]
- m_{3/2} \mu_{ij}\Big\}
\\ &
+
\left(2m_{3/2}^2+\frac{V}{M^2_{\rm Pl}}\right) {Z}_{ij} -    
m_{3/2}{\overline{F}}^{\overline{m}} 
\partial_{\overline{m}} Z_{ij} 
\\ &
+
m_{3/2} F^m \left[ \partial_m Z_{i j} - {\tilde K^{{ \delta} {\overline{\rho}} }}\left(
\partial_m {\tilde K_{{\overline{\rho}}{ i}}} Z_{\delta j}
+
\partial_m {\tilde K_{{\overline{\rho}}{ j}}} Z_{\delta i}\right)\right]
\\ &
-\ {\overline{F}}^{\overline{m}} F^n \left[ \partial_{\overline{m}}
\partial_n 
Z_{ij} - {\tilde K^{{ \delta} {\overline{\rho}} }}\left(
\partial_n {\tilde K_{{\overline{\rho}}{ i}}} \partial_{\overline{m}} 
Z_{\delta j}
+
\partial_n {\tilde K_{{\overline{\rho}}{ j}}} \partial_{\overline{m}} 
Z_{\delta i}\right)\right].
\eeq

We can use these expressions to derive the contribution (\ref{maratio}) to the axion mass due to the addition of an R-breaking constant in the superpotential as discussed in Section \ref{rexplicitbreak}. Given the form of $F^m$ in  eq.~(\ref{Fterm}) the explicit R-breaking constant in the superpotential, $W_0$,  will result in contributions to the axion mass by inducing explicit R-breaking $A$ and $B$ (and potentially higher order) terms for the PQ charged scalars. Note that without such a $W_0$ the A and B terms would instead at most spontaneously break R. The exact form of these $A$ and $B$ terms is highly model dependent, especially in the case of a non trivial K\"ahler potential $\tilde{K}_{\overline{i}j},Z_{ij} \neq 0$. 

Setting U(1)${}_{\rm PQ}$ = U(1)${}_R$ greatly restricts the allowed form of the super- and K\"ahler potential in eq.~(\ref{W}), forbidding many terms in eqs.~(\ref{Bterm}) \& (\ref{Aterm}). However in PQ axion constructions, like for example the KSVZ and DFSZ models, the superpotential will generically include some polynomial in the PQ charged superfields $W \supset \mathcal{O}(X,Y, H_{u},H_{d},\cdots)$ in order to generate the appropriate stabilising potential. Thus in order to implement the PQ mechanism at least one of $\mu_{\alpha\beta}$, $Y_{\alpha\beta\gamma}$, or a similar coefficient corresponding to a higher dimensional operator in the superpotential, needs to be nonzero. Hence the corresponding contributions to the axion mass due to the R-breaking constant in the superpotential are generic for realistic implementations of the PQ mechanism. The size of this mass contribution can be estimated by arguments similar to those in Section \ref{se3.4}. For example, taking the polynominal $W\supset \frac{1}{M^{n+m-3}_{\rm Pl}}X^nY^m$ and plugging in VEVs $\langle X\rangle \sim \langle Y\rangle\sim f_a$  (from eq.~(\ref{<X>})) into the first line of eq.~(\ref{Bterm}) we find a mass contribution 
\beq \label{breakmassw}
(m_a^{\not R})^2 \sim \left(\frac{f_a}{M_{\rm pl}}\right)^{n+m}\frac{M^3_{\rm Pl}m_{3/2}}{f^2_a}.
\eeq
For concreteness, taking the Kim-Nilles superpotential eq.~(\ref{Hitoshi}) and the gauge mediation scenario $\Lambda \sim M_{\rm Pl}$ with $m_{\rm soft} \sim m_{3/2}$ we arrive at $m_a^{\not R} \sim m_{3/2}$. As discussed in Section \ref{se3.4} there will in general also be contributions from the K\"ahler terms. From the third line of eq.~(\ref{Bterm}) on has $B_{ij} \supset 2m^2_{3/2}Z_{ij}$
and we find  generic contribution to the axion mass from the explicit R-breaking due to $W_0\neq0$ to be
  \beq \label{breakmass}
(m_a^{\not R})^2\sim 2m^2_{3/2}Z_{ij}~,
\eeq
where $Z_{ij}\sim (\langle z\rangle / M_{\rm Pl})^d$, corresponding to the dimension of the leading R-invariant operator.

If the R-charges are such that only high dimensional breaking operators are induced (for similar reasons as discussed in Section~\ref{se3.4}), the mass ratio eq.~(\ref{breakmass}) is suppressed by  appropriate factors of $f_a/M_{\rm Pl}$.  We use eqs.~(\ref{breakmassw}) \& (\ref{breakmass}) in the estimate in Section \ref{rexplicitbreak}.

\subsection{R-symmetries and Fayet-Iliopoulos Terms} \label{superconformal}

As discussed in Section \ref{sec:5}, global symmetries can be gauged in the UV via the Green Schwarz mechanism, which potentially provides an elegant manner of avoiding the axion quality problem \cite{Kamionkowski:1992mf,Holman:1992us,Barr:1992qq}. Since we are identifying U(1)${}_R$ with the PQ symmetry this amounts to gauging the R-symmetry. Notably, R-symmetries cannot be gauged in global supersymmetry, as can be seen by the fact that under a local R-symmetry $Q_{\alpha} \rightarrow e^{-i\epsilon(x)}Q_{\alpha}$. Thus, the R-symmetry can only be consistently gauged if the supersymmetry generators are also position dependent, and hence the gauging of R-symmetry is only possible in supergravity. 
Moreover, in supergravity the gauging of U(1)${}_R$ is tied to the presence of constant Fayet-Iliopoulos (FI) terms
\cite{Stelle:1978wj,Barbieri:1982ac,Freedman:1976uk}.
The consistent coupling of a FI term to a U(1) vector multiplet in supergravity generally (with some caveats \cite{Cribiori:2017laj}) has the effect of gauging an R-symmetry and vice-versa.

For a given gauge symmetry, it is not restrictive to assume that the K\"ahler potential is gauge invariant. The superpotential is either gauge invariant or, in the case of a $U(1)_R$ symmetry, is rotated by a real phase. Hence
\beq\label{A4}
D\supset iM^2_{\rm Pl}\frac{W_{j}}{W}X^{j}_{a} \equiv \xi_{a}.
\eeq
The constant contributions $\xi_{a}$ to the Killing potential are referred to as Fayet-Iliopoulos terms, in accord with global supersymmetry. 
Notably, under a local U(1)${}_R$ transformation we expect the superpotential to transform as 
\beq \label{fi}
\delta W= W_{j} \delta \phi^j = W_{j}X^{j}_{R} \epsilon_{R}(x) =i\xi\epsilon_{R}(x)W ~,
\eeq
where in the last step we used that the definition of an R-symmetry. Thus we find that a U(1)${}_R$ symmetry under which the superpotential has charge $\xi$ leads to a constant FI term  contribution to the Killing potential
\beq\label{kapparel}
D\supset \xi_{R}=-\xi M_{\rm Pl}^2.
\eeq
The converse statement is also true: the consistent coupling of a constant FI term in supergravity generally requires the superpotential to transform non trivially and hence the symmetry gauged by the vector multiplet is an R-symmetry. This converse statement is non-trivial and is laid out in the remainder of this section.
 The connection between constant FI terms and gauged R-symmetries is most readily seen in the superconformal formalism for supergravity.

In \cite{Stelle:1978wj,Barbieri:1982ac} the action for the Fayet-Illiopoulos term in supergravity in the superconformal framework including matter chiral superfields $\Phi^i$ with a general superpotential $W$ was derived. 
Here we collect together some of the useful general theory, very closely emulating the Appendix of \cite{Antoniadis:2014iea}.
Choosing the normalisation such that the R-symmetry transforms the superpotental as $W(\Phi^i)\rightarrow e^{+ i\xi \Lambda}W(\Phi^i)$, we consider the  action below
\beq \label{sc_action}
\int
d^4x\,&
\Big\{
 d^4\vartheta\,\mathbf{E}\,
\Big[  
(-3M^2_{\rm Pl})\, S_0^\dagger \,e^{2 \,(\xi/3)\,V_R} S_0\,e^{-M^{-2}_{\rm Pl}\, K_0/3}\Big]\\
&+
\int d^2\vartheta \mathcal{E}\Big[
 S_0^3\,W(\Phi^i)+\frac{1}{4}\mathcal{F}_{ab}{\rm Tr}~{\mathcal{W}^{\alpha,a}\mathcal{W}^b_{\alpha}}+
{\rm h.c.}\Big]\Big\}~,
\eeq
where $S_0$ is the conformal compensator, $K_0$ is the K\"ahler potential of the superfields, $\mathbf{E}$ is the superspace measure, $\mathcal{E}$ is the chiral superspace measure, $V_R$ is the vector superfield gauging U(1)${}_R$, and $\xi$ is the constant Fayet-Iliopoulos term in the sense of eq.~(\ref{fi}). 
Thus $\xi_R$ is the FI term appearing in the U(1)${}_R$ D-term and $\xi$ is the charge of the superpotential, and we next look to establish a connection between the two.
The R-charges of the chiral superfields are defined such that they transform as $\Phi^{j} \rightarrow e^{iq_{j}\Lambda}\Phi^{j}$ under the R-symmetry.

The fact that eq.~(\ref{sc_action}) corresponds to a constant FI term is evident in the $M_{\rm Pl} \rightarrow \infty$ limit, in which case
\beq
\mathcal{L}\supset \int d^4\vartheta \,(K_0-2M^2_{\rm Pl}\xi V_R)~.
\eeq
If we consider the case $K_0\supset \Phi^{\dagger} \exp(-2qV_{R})\Phi$ one obtains
\beq
\mathcal{L}\supset -q\vert \phi\vert^2-\xi M^2_{\rm Pl}\,D+D^2/2~,
\eeq
with the familiar $D$-term expression
\beq
D^2\sim \left(q\vert\phi\vert^2+\xi M^2_{\rm Pl}\right)^2~.
\eeq
The Lagrangian eq.~(\ref{sc_action}) is invariant under the following U(1) transformation
\beq\label{B5}
V_R & \rightarrow   V_R+\frac{i}{2}(\Lambda-\Lambda^\dagger),\\
 \overline D\Lambda &=0~,\\
S_0 &\rightarrow 
e^{ - i \,\xi/3\,\Lambda}\,S_0~,\\
W&\rightarrow 
e^{+ i\,\xi \Lambda}W~.
\eeq
This U(1) transformation requires the superpotential to transforms non-trivially and thus is an R-symmetry with $R[W]=\xi$.

An issue arises in that the conformal compensator $S_0$ must also transform implying that the symmetry is not well defined. The transformation of $S_0$ can be compensated with a choice of super-Weyl gauge in order to keep manifest supersymmetry and holomorphicity.
Taking the super-Weyl gauge to be
$S_0=s_0+\vartheta^2\,F$ makes the conformal compensator neutral under U(1)${}_R$. This can be achieved by combining the U(1)${}_R$ transformation with a super-Weyl transformation. The action is invariant under the following super-Weyl symmetry ($\tau \in \mathbb{C}$)
\beq \label{weyl}
\lambda &\rightarrow e^{-3\tau}\lambda,
\hspace{27mm}
\mathbf{E}\rightarrow e^{2\tau +2\overline\tau}\,\mathbf{E}, \\
 \mathcal{E} &\rightarrow e^{6\,\tau}\,\mathcal{E},
\hspace{25mm}
\mathcal{W}_\alpha\rightarrow e^{-3\tau}\,\mathcal{W}_\alpha,\\
V^{(a)} &\rightarrow V^{(a)},
\hspace{28mm}
S_0\rightarrow e^{-2\,\tau}\,S_0,\\
W &\rightarrow W,
\hspace{29mm}
\overline{\mathcal{W}}^{\dot\alpha} 
\rightarrow 
e^{-3\overline\tau}
\overline{\mathcal{W}}^{\dot\alpha}~.
\eeq
Comparing eqns.~(\ref{B5}) \& (\ref{weyl}) it can be seen that the conformal compensator $S_0$ is neutral under the combined transformation if
\beq \label{tau}
\tau=-\frac{i\,\xi \Lambda}{6}~.
\eeq
Combining eq.~(\ref{weyl}) \& (\ref{tau}) leads to relations between the R-charges of various structures. 
For example, from the super Weyl transformation of $\mathcal{W_{\alpha}}$ it follows that $R[\lambda]=R[\mathcal{W}]=\xi/2$. 
Similarly, since $R[V]=0$ and $V\supset \overline{\vartheta}\overline{\vartheta}\vartheta\lambda$ this implies that $R[\vartheta]=R[\lambda]=\xi/2$. 

Finally, by considering a linear R-transform $X^{j}=iq^{j}\phi^{j}$, the R-charge of the fermion component $\psi^j$ of a chiral superfield $\Phi^j$ with R-charge $q_i$ is found to be
\beq \label{fcharge}
\mathcal{L}\supset X^{j}_{R}\overline\psi^{j}\,\overline\lambda+{\rm h.c.} \quad \Rightarrow\quad R[\psi^{j}]= q_j - \xi/2~.
\eeq
The R-charge of the gauginos, and the difference in R-charge between fermionic and bosonic components of chiral superfields, is set by the choice of $\xi$.

Similarly the R-charge of the gravitino $\psi_{\mu}$ can be identified by considering, for example,
\beq
\mathcal{L}\supset
M^{-2}_{\rm Pl}e^{M^{-2}_{\rm Pl}K/2}W\overline\psi_{\mu}P_{R}\gamma^{\mu\nu}\overline\psi_{\nu}+{\rm h.c.} 
\eeq
from which it follows that
\beq
R[{\psi_{3/2}}]=\frac{1}{2}\,R[W]=\frac{\xi}{2}.
\eeq
For $\xi=2$ one finds the canonical global supersymmetry assignments: $R[\vartheta]=1$ and $R[W]=2$.

\section{Closed String Axions}
\label{app-string}
In this appendix we will briefly introduce string axions in order to better motivate the Green Schwarz mechanism.  For brevity we will discuss only the string-independent axion below, following \cite{Choi:2014uaa}. See e.g.~\cite{Svrcek:2006yi} for more complete discussions.

String theory contains a variety of  higher-dimensional objects, $p-1$ branes, which couple to antisymmetric $p$-form gauge fields 
$C_p$, which transform under a gauge symmetry
\beq \label{pform}
C_p \rightarrow C_p + d\Lambda_{p-1}~,
\eeq
where $\Lambda_{p-1}$ is
a $(p-1)$-form gauge parameter.
Upon compactifying the theory to 4-dimensions with an internal space involving a $p$-cycle $\Sigma_p$, the resulting 4D effective theory zero mode expansion of the gauge field contains an axion-like field $\theta_{\rm s}$
\beq
C_p(x,y) = \theta_{\rm s}(x) \omega_p(y),
\eeq
where $x$ and $y$ denote the 4D and internal space coordinates, respectively, and
 $\omega_p$ is a harmonic $p$-form dual to $\Sigma_p$ with $\int_{\Sigma_p} \omega_p =1$.
Since $\omega_p(y) = d\Omega_{p-1}(y)$ is  locally an exact $p$-form, the gauge symmetry of eq.~(\ref{pform}) is locally equivalent to a shift symmetry of the axion U$(1)_{\rm cl}$ given by
\beq \label{shift}
{\rm U}(1)_{\rm cl}: \,\, \theta_{\rm s}(x) \rightarrow \theta_{\rm s}(x)+{\rm constant}~
\eeq
while being globally ill-defined.
For compatifications that preserve 4D $\mathcal{N} = 1 $ supergravity the above axion  corresponds to the imaginary part for the scalar component of the axion-dilaton multiplet $T=\tau + i\theta_{\rm s}$.

The shift symmetry of eq.~(\ref{shift}) can be explicitly broken by stringy instantons 
wrapping $\Sigma_p$, as well by coupling of the axion to the Standard Model gauge instantons. The coupling of the axion to the Standard Model instantons is generated by
\beq \label{axioncouple}
\int C_p\wedge G\wedge G
\quad \rightarrow \quad \int_{\mathcal{M}_4} \theta_{\rm s}G\wedge G \int_{\Sigma_p} \omega_p.
\eeq
This coupling allows for the stringy axion $\theta_{\rm s}$ to act as the QCD axion, potentially solving the strong CP problem. We will assume that the stringy instanton contribution is small enough so as to avoid quality problems, string models achieving this can be found in e.g.~\cite{Demirtas:2021gsq}.  

The closed string axion decay constant is derived from the 4D effective action of the axion-dilaton superfield. Given the K\"ahler potential $K$ and the gauge kinetic function
$\mathcal{F}_{ab}$ 
\beq
K &=-M_{\rm Pl}^{2}K_0(T+\overline{T}), \\
 \mathcal{F}_{ab} &= \delta_{ab} k_{a}T + \cdots
\eeq
where $k_{a}$ is the Kac-Moody level of a given gauge group labeled by $a$. For a given gauge kinetic function $\mathcal{F}_{ab}$ one obtains the Lagrangian
\beq \label{gaugekin}
\mathcal{L} \supset -\frac{1}{4}{\rm Re}[\mathcal{F}_{ab}] G^{a}G^{b} + \frac{1}{4}{\rm Im}[\mathcal{F}_{ab}] \tilde G^{a}G^{b}~.
\eeq

The kinetic term for the scalar component $T^{s}$ of the dilaton superfield (in terms of the K\"ahler potential) is given by
\beq \label{kinterm}
\mathcal{L} \supset -K_{T\overline{T}}D_{\mu}T^{s}D^{\mu}\overline{T^{s}}~.
\eeq
The relevant part of the Lagrangian for  $\theta_{\rm s}$ is thus
\beq
\mathcal{L} \supset 
K_{T\overline{T}}\partial_\mu \theta_{\rm s}
\partial^\mu \theta_{\rm s}
+ \frac{1}{4} \theta_{\rm s} k_{a} G^{a\mu\nu}\tilde G^{a}_{\mu\nu}~.
\eeq
Performing a field redefinition to write the Lagrangian in terms of a canonically normalised string axion $a_{\rm st}$ yields
\beq
\mathcal{L} \supset 
=\frac{1}{2}\partial_\mu a_{\rm st} \partial^\mu a_{\rm st}
+ \frac{1}{32\pi^2}\frac{a_{\rm st}}{f_{\rm s}}k_{A}G^{A\mu\nu}\tilde G^A_{\mu\nu}+ \cdots
\eeq
where the decay constant for this string axion can be identified as
\beq \label{fcl}
f_{\rm s} = \frac{1}{8\pi^2}\sqrt{2K_{T\overline T}}~.
\eeq

Furthermore, the VEV of the modulus field $T$ sets the 
string coupling constant
\beq \label{vevdilation}
 1/g_s^2 = {\rm Re}[\mathcal{F}_{ab}(T)]  = \langle\tau\rangle~.
\eeq
The GUT coupling and the string coupling $g_s$ are connected, for instance, in heterotic string theory  one finds that $g^{-2}_{\rm GUT}=\mathcal{V}g^{-2}_{s}= \langle\tau\rangle$, where $\mathcal{V}$ is the compactification volume. 
To obtain an order unity coupling constant one  requires $|\langle T \rangle| \sim \mathcal{O}(1)$. Since the K\"ahler metric for the modulus is generally $K_{T\overline T} \sim \mathcal{O}(M^2_{\rm Pl})$ it follows that  
\beq
f_{\rm s} \sim \frac{M_{\rm Pl}}{8\pi^2}~.
\eeq
This decay constant is commonly thought to be too large to be the QCD axion  \cite{Svrcek:2006yi,Choi:2014uaa}.

More desirable values for the decay constant may be obtained by a mixing of the closed string axion with open string axions arising from the gauging of an anomalous U(1)${}_A$ symmetry via the GS mechanism \cite{Green:1984sg}, as we discuss in Section \ref{sec:5}. In what follows we derive the effective Lagrangian used as a starting point in that section.
As stated in the main text, gauge anomalies can be cancelled via the GS mechanism by allowing U(1)${}_A$ to act on the modulus $T$ as a shift symmetry as follows
 \beq \label{define2}
{\rm U}(1)_A: \qquad V_A &\rightarrow V_A+\frac{i}{2}(\Lambda- \overline{\Lambda}), \\
T&\rightarrow T+i\delta_{\rm GS}\Lambda, \\
\phi_j& \rightarrow e^{iq_j\Lambda}\phi_j~.
\eeq
The coupling of eq.~(\ref{axioncouple}) allows the gauge anomalies of the U(1)${}_A$ symmetry to be cancelled by the shift of the closed string axion $\theta_{\rm s}$.

Consider the anomalous U(1)${}_A$ of eq.~(\ref{define2}) with some chiral superfields $\phi_j$ charged under U(1)${}_A$ with charges $q_{j}$ and the following form of the K\"ahler potential and gauge kinetic function 
\beq
K = -M^{2}_{\rm Pl} K_0&(T+\overline{T}-2\delta_{\rm GS}V_A)\\
& + Z_j(T+\overline{T}-2\delta_{\rm GS}V_A)
\phi_j^* e^{-2q_j V_A} \phi_j
+ \cdots\\[5pt]
 \label{kinfunc}
\mathcal{F}_{ab} = \delta_{ab}k_{a}T +& \cdots
\eeq
To simplify our discussion we fix $Z_{j}(..)=1$ (see \cite{Kim:2008kw} for the general case), 
and take \cite{Ahn:2016typ,Buchbinder:2014qca}
\beq K_{0} = {\rm ln}(T+\overline{T}-2\delta_{\rm GS}V_A) .
\eeq
Using eqns.~(\ref{DDD2}) \& (\ref{gaugekin})-(\ref{kinterm}), 
the relevant part of the Lagrangian is given by 
\beq 
{\cal L} =&-\frac{1}{4g^{2}_{a}}G^{a,\mu\nu}G^{a}_{\mu\nu}+ K_{T\overline{T}}\left(\partial_\mu \theta_{\rm s}- \delta_{\rm GS}A_\mu\right)^2
+ D_\mu \phi^*_{j} D^\mu \phi_{j}\\
&+\frac{g^{2}_R}{2}\left(-\delta_{\rm GS} K_{T} -\sum_{j}q_{j}|\phi_{j}|^2 +\xi_{R}\right)^2\\
&+\frac{1}{4} k_{a}\theta_{\rm s} G^{a\mu\nu}\tilde G^{a}_{\mu\nu}+{\cal L}_{\rm Matter}+\cdots
\eeq
This is the form of eq.~(\ref{fullL}) stated in Section \ref{sec:5} as our starting point for discussing the Green Schwarz mechanism.


\end{document}